\documentclass{jfm}
\usepackage{caption}
\usepackage{newtxtext}
\usepackage{newtxmath}
\usepackage{bm}
\usepackage[authoryear]{natbib} 
\usepackage{hyperref}
\usepackage{graphicx}
\usepackage{tikz}
\usepackage{subcaption}
\usepackage{longtable}

\usepackage{multicol} 
\usepackage{tabularx}
\usepackage{booktabs}
\usepackage{subcaption}

\hypersetup{
    colorlinks = true,
    urlcolor   = blue,
    citecolor  = black,
}

\newcommand{\RomanNumeralCaps}[1]
\linenumbers

\usepackage[dvipsnames]{xcolor}
\newcommand{\CRCA}[1]{\textcolor{black}{#1}}
\newcommand{\SHU}[1]{\textcolor{black}{#1}}

\title{Pinch-off dynamics for a Newtonian liquid thread draining in a viscoplastic medium}

\author[S. Yang, K. Zinelis, S. Chakraborty, C. R. Constante-Amores]
{S. Yang$^{1}$, K. Zinelis$^{2}$, S. Chakraborty$^{3}$, 
  C. R. Constante-Amores$^{1}$\corresp{\email{crconsta@illinois.edu}}}

\affiliation{
  $^{1}$Department of Mechanical Science and Engineering, University of Illinois Urbana-Champaign\\
  $^{2}$Department of Chemical Engineering, Massachusetts Institute of Technology\\
  $^{3}$Department of Chemical \& Biomolecular Engineering, Johns Hopkins University
}

\begin{document}
\maketitle

\begin{abstract}
The pinch-off dynamics of a Newtonian liquid thread embedded in a viscoplastic medium is investigated using direct numerical simulations and theory. Thread breakup is controlled by a rupture/arrest threshold in the plastocapillary number, $\mathcal{J}$, set primarily by the balance between capillary and yield stresses.
\textcolor{black}{For a fixed initial perturbation amplitude and wavenumber, this threshold is largely independent of the viscosity ratio at low Ohnesorge numbers. However, $\mathcal{J}$ 
decreases for smaller imposed perturbation amplitudes, and also decreases as the Ohnesorge number is increased, because viscous stresses provide additional damping of the capillary-driven motion.} 
Regime maps in the Ohnesorge–plastocapillary number plane reveal distinct boundaries \CRCA{ between pinch-off and arrested states. After the thread begins to thin, we derive reduced temporal evolution equations for the minimum thread radius. These equations capture the subsequent thinning dynamics in excellent agreement with the simulations. }
These results provide a predictive framework  for thread dynamics in
{viscoplastic}  fluids.
The obtained insights are relevant for embedded additive manufacturing involving fluid threads in {viscoplastic} media.
\end{abstract}

\section{Introduction}

The thinning and pinch-off of liquid threads is a canonical problem in free-surface flows \citep{eggers_physics_2008}, with applications spanning inkjet printing \citep{derby_inkjet_2010,rump_role_2024}, electrohydrodynamic spraying \citep{lopez-herrera_one-dimensional_1999}, and fiber or additive manufacturing \citep{sun_overview_2022,bankapalli_filament_2023}. In Newtonian fluids, capillary, viscous, and inertial forces govern the growth of perturbations along a cylindrical thread, giving rise to the classical Plateau--Rayleigh instability and nonlinear pinch-off when perturbations exceed a critical wavelength \citep{rayleigh_xvi_1892,basaran_nonstandard_2013,castrejon-pita_breakup_2012,stone_dynamics_1994}. 



In Newtonian fluid, this complex capillary instability that can ultimately result in a non-linear pinch-off is governed by capillary force, viscosity, and inertia \citep{castrejon-pita_breakup_2012, notz_dynamics_2004,stone_dynamics_1994, eggers_instability_2014}. Research on this model initiated by the pioneering work of \citet{PlateauExperimentalAT}, who demonstrated that cylindrical liquid threads become unstable when subjected to perturbations with adequate wavelengths. This capillary instability arises because the initial reduction in surface area {results} in a drop in surface energy that drives the growth of perturbations \citep{eggers_physics_2008}. Such instability is called Plateau-Rayleigh instability. Following this work, subsequent studies by \citet{608083d4-5f98-306f-8765-fe627d44a54d, rayleigh_xvi_1892} conducted stability analysis and identified the wavelength of maximum instability \citep{basaran_nonstandard_2013, eggers_physics_2008}.

Three scaling law regimes have been established  to describe thread pinch-off dynamics. \citet{eggers_universal_1993} first established inertial-viscous scaling laws for thread dynamics, demonstrating that the flows display self-similar behavior characterized by scaling exponents for both neck shape and velocity field. \citet{papageorgiou_breakup_1995} subsequently developed viscous scaling laws where capillary and viscous forces dominate, later extended by \citet{lister_capillary_1998} to include external phase viscosity effects. Finally, the inertial regime addresses inviscid thread breakup governed by inertial and capillary forces \citep{day_self-similar_1998, wee_breakup_2022}.

Later studies on this paradigm also studied the effect of surfactants \citep{craster_pinchoff_2002,xu_can_2007, kamat_role_2018} and incorporated viscoelasticity within the thread \citep{eggers_instability_2014,Deike_2018}. By contrast, the influence of a viscoplastic surrounding medium, a non-Newtonian material exhibiting yield-stress behavior, remains largely unexplored. This is because yield stress fluids  behave like elastic solids under stresses below a critical yield stress,  and flow like non-Newtonian fluids (e.g., shear-thinning or shear-thickening) when this stress is exceeded \citep{Balmforth_Hewitt_2025,hossain_critical_2025,sanjay_bursting_2021}.  This dual behavior raises fundamental questions about how capillary, viscous, and yield
stresses interact to determine thread rupture.

{ In geometries beyond liquid threads, viscoplastic effects have also received considerable attention. A seminal contribution is that of \citet{beris_creeping_1985}, who examined the creeping motion of a rigid sphere in a Bingham fluid and demonstrated the existence of a critical yield number beyond which the sphere remains arrested. Their work established how the balance between net gravitational force and material yield stress determines whether motion occurs. Subsequent studies extended this framework to deformable interfaces, exploring the interplay between buoyancy, capillarity and yield stress in the rise of bubbles and droplets, as well as in interfacial rupture and stability problems \citep{Pourzahedi_Chaparian_Roustaei_Frigaard_2022, Pourzahedi_2024, sanjay_bursting_2021,deka}.}

{Recent experiments by \citet{hossain_critical_2025} demonstrated that a Newtonian filament embedded within a carbopol  bath can remain stable below a critical plastocapillary threshold, identifying a universal dimensionless criterion based on the balance between capillary and yield stresses. Although carbopol suspensions may exhibit weak viscoelastic effects, the stability transition was attributed primarily to the presence of a finite yield stress. In the present study, we isolate the role of yield stress by modelling the surrounding medium as an idealised viscoplastic fluid (i.e., not viscoelastic effects) and follow the fully nonlinear evolution of a perturbed, infinitely long Newtonian filament to determine the transition between arrested and rupturing regimes.}

{Our results reveal a consistent rupture threshold}, primarily set by the balance of capillary and yield stresses and largely independent of viscosity ratio at low Ohnesorge numbers ($\mathcal{O}h$, comparing the inertial-capillary to inertial-viscous timescales). We present regime maps in terms of $\mathcal{O}h$ and the plastocapillary number $\mathcal{J}$ (which accounts for the competition between the capillary and yield stresses), and identify scaling laws for the minimum thread radius and maximum axial velocity over time. {These results provide fundamental insights into thread dynamics in viscoplastic media.} This paper is organized as follows: we briefly describe the numerical method, then present a regime map illustrating the influence of viscoplasticity on the dynamics, along with new scaling laws for the maximum axial velocity and minimum thread radius over time.

\section{Numerical formulation}

\subsection{{Governing equations and numerical method}}

We solve the axisymmetric, two-phase, incompressible Navier--Stokes equations using the open-source Basilisk library \citep{popinet_accurate_2009,popinet2013basilisk}.
{All dimensionless variables are  represented by tildes.} Distances, time, and pressures are nondimensionalized by $R_0$, $t_i = \sqrt{{\rho_l} R_0^3 /\sigma}$, and $\sigma/R_0$, respectively, where $\sigma$ and {$\rho_l$} represent the surface tension and thread density, and $R_0$ is the radius of the undeformed thread. Here $t_i$ represents the inertial-capillary time scale. As a result of this scaling, the dimensionless equations read

\begin{equation}
\tilde{\nabla}\cdot \tilde{\mathbf{u}} = 0, \quad 
{ \tilde{\rho}}
\left(
\frac{\partial \tilde{\mathbf{u}}}{\partial \tilde{t}}
+\tilde{\mathbf{u}}\cdot \tilde{\nabla}\tilde{\mathbf{u}}
\right)
=-\tilde{\nabla}\tilde{p}
+\tilde{\nabla}\cdot \tilde{\mathcal{T}}+
{\tilde{\boldsymbol{f}}_\sigma},
\label{eq:ns_dimless}
\end{equation}

\noindent where dimensionless parameters $\tilde{\mathbf{u}}$ is the velocity, $\tilde{p}$ is the pressure, $\tilde{\boldsymbol{\mathcal{T}}}$ is the deviatoric stress tensor, {$\tilde{\boldsymbol{f}}_\sigma$ is the surface tension force. }

{
In a one-fluid formulation,  the deviatoric stress depends on the local material properties. The thread (inner) phase is Newtonian with constant viscosity $\mu_l$ and no yield stress.
The external phase is modeled as a Herschel--Bulkley fluid, whose deviatoric stress tensor
\begin{equation}
\tilde{\boldsymbol{\mathcal{T}}}
=
2\,\tilde{\mu}_{\mathrm{eff}}(\|\tilde{\boldsymbol{\mathcal{D}}}\|)\,\tilde{\boldsymbol{\mathcal{D}}},
\label{eq:HB_stress}
\end{equation}
where the dimensionless rate-of-deformation tensor,$\tilde{\boldsymbol{\mathcal{D}}},$  is defined as
\begin{equation}
\tilde{\boldsymbol{\mathcal{D}}}
=
\frac{1}{2}
\left(
\tilde{\nabla}\tilde{\mathbf{u}}
+
(\tilde{\nabla}\tilde{\mathbf{u}})^{T}
\right),
\qquad
\|\tilde{\boldsymbol{\mathcal{D}}}\|
=
\sqrt{\textcolor{black}{\frac{1}{2}}\,\tilde{\boldsymbol{\mathcal{D}}}:\tilde{\boldsymbol{\mathcal{D}}}}.
\end{equation}
The dimensionless apparent viscosity, $\tilde{\mu}_{\mathrm{eff}},$ for the Herschel--Bulkley fluid is given by
\begin{equation}
\tilde{\mu}_{\mathrm{eff}} =
\frac{\mathcal{J}}{2\|\tilde{\boldsymbol{\mathcal{D}}}\|+\tilde{\varepsilon}}
+
m\mathcal{O}h\left(2\|\tilde{\boldsymbol{\mathcal{D}}}\|+\tilde{\varepsilon}\right)^{\,n-1},
\label{eq:HB_viscosity}
\end{equation}}


The dimensionless numbers controlling the flow are the plastocapillary number $\mathcal{J} = \mathcal{T}_y R_0 / \sigma$, the Ohnesorge number $\mathcal{O}h = \mu_l / \sqrt{\rho_l \sigma R_0}$, the power-law index $n$, the normalized viscosity ratio $m = K_b \dot{\gamma}_c^{n-1} / \mu_l$, and the density ratio $\rho_r = \rho_b / \rho_l$, where $\dot{\gamma}_c$ denotes the characteristic shear rate, $\rho_b$ is external phase density, {$K_b$ is the consistency index of the external phase, $\mu_l$ is thread viscosity, and $\tilde{\varepsilon}$ is a small regularization parameter introduced for numerical stability.}
{
In the one-fluid formulation, material properties are expressed as
volume-fraction-weighted fields. The dimensionless density and viscosity are
$\tilde{\rho} = f + (1-f)\rho^{*},$ and
$\tilde{\mu} = f + (1-f)\mu^{*}$,
where $\rho^{*}=\rho_r$ and $\mu^{*}=\mu_{\mathrm{eff}}/\mu_{l}$.
}

{
Surface tension is modeled in Basilisk using a one-fluid VOF/CSF formulation. The interface is tracked by the dimensionless volume fraction $f$, and the unit normal is $\tilde{\mathbf{n}}=\tilde{\nabla} f/\|\tilde{\nabla} f\|$. The curvature is defined as $\tilde{\kappa}=\tilde{\nabla}\cdot\tilde{\mathbf{n}}$, which corresponds to the sum of principal curvatures (i.e. $\tilde{\kappa}=2\tilde{H}$). The capillary force enters the (dimensionless) momentum equation as $\tilde{\boldsymbol{f}}_\sigma=\tilde{\sigma}\,\tilde{\kappa}\,\tilde{\delta}_\Gamma\,\tilde{\mathbf{n}}$, where $\tilde{\delta}_\Gamma$ is a regularized surface delta function supported in the interfacial cells. Curvature is computed using Basilisk’s height-function method. For more numerical details about surface tension forces modelling in Basilisk, we direct the reader to \citet{popinet_accurate_2009,popinet2018numerical}. Tildes are dropped henceforth.}

\subsection{{Numerical set-up}}

\begin{figure}
\centering
\begin{tabular}{c}
\includegraphics[width=0.75\linewidth]{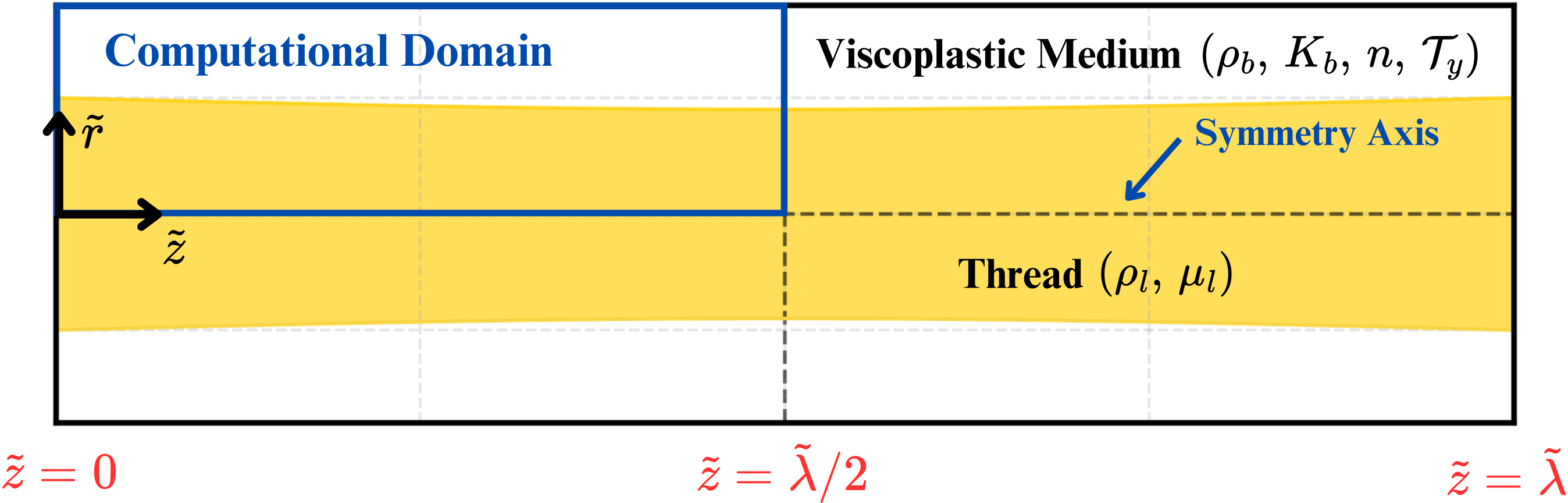}\\
(a)\\
\includegraphics[width=0.7\linewidth]{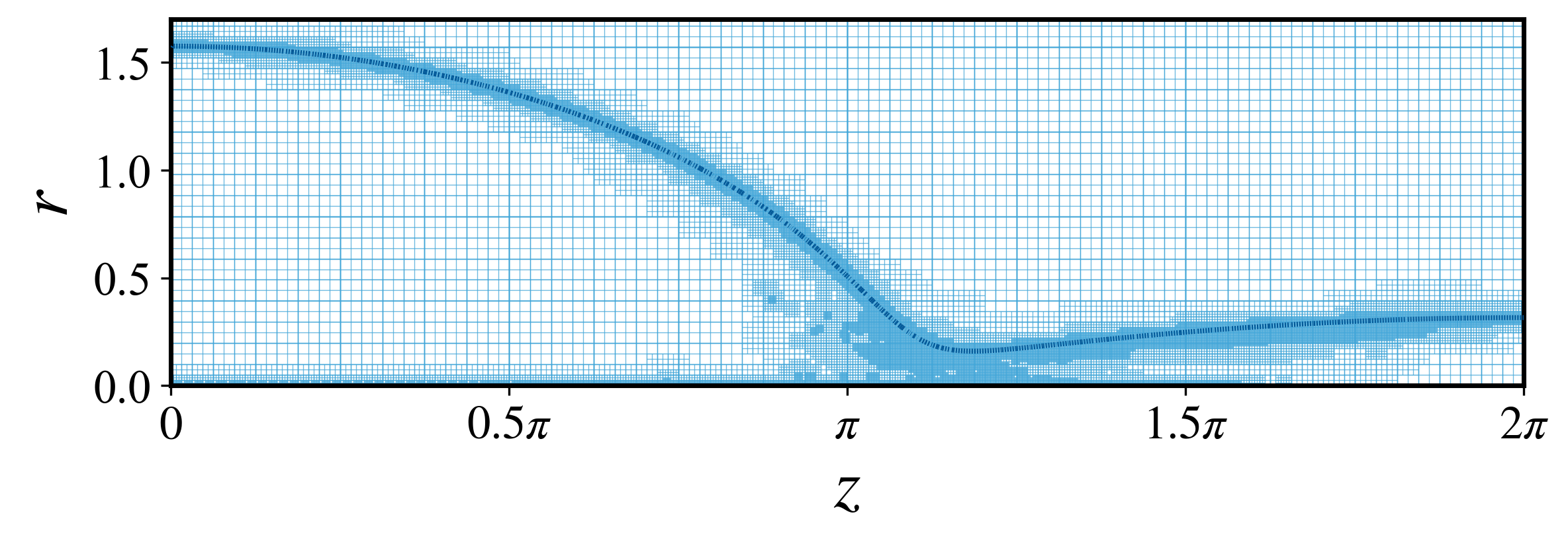}\\
(b)\\
\end{tabular}
\caption{(a) Sketch of the numerical configuration. (b) Representative mesh for a simulation exhibiting a capillary singularity.}
 \label{fig:geometry}
\end{figure}

\noindent
The initial numerical setup follows closely the work of \citet{castrejon2015plethora} and  \citet{Deike_2018}.  We consider a thread of initial radius $R_0$ surrounded by viscoplastic fluid in absence of gravity. {Figure \ref{fig:geometry} shows a schematic of the numerical configuration, highlighting the computational domain. It also includes a snapshot of the interface near the capillary singularity together with the corresponding computational mesh, which is highly refined in the vicinity of the interface}. An initial perturbation, $\delta$ is applied to initiate the capillary thinning of the filament, so that the initial filament profile, $\tilde{h}(z,0)$, is given by
\textcolor{black}{
$\tilde{h}(\tilde{z},0)/R_0 = 1 + \delta \cos \left(k\tilde{z}\right)$,
where $\tilde{z}$ is the nondimensional axial coordinate, $k=2\pi/\tilde{\lambda}$
is the nondimensional wavenumber, and the axial extent of the domain is
$0 \le \tilde{z} \le \tilde{\lambda}/2=2\pi$, corresponding to $k=1/2$.} 
The perturbation wavelength and amplitude followed standard practice in works by \citet{Deike_2018, wee_breakup_2022, constante-amores_dynamics_2020,kamat_role_2018}.
%
To robustly trigger yielding across the parameter space, the perturbation amplitude was fixed at $\delta = 0.1$ throughout the study.
\textcolor{black}{Section \ref{sec: geom_Jc} present the  role of both the perturbation amplitude and the length of the thread on the dynamics. }
The bottom boundary corresponds to the axis of symmetry, while the lateral boundaries are treated as symmetry planes.
{Along the axis of symmetry, axisymmetric regularity is enforced by setting $u_r = 0$, $u_\theta = 0$, and $\partial p / \partial r = 0$.}
{The flow is assumed incompressible and isothermal, and inertial, viscous, and capillary effects are retained. Gravity and external forcing are neglected, and the surrounding fluid is quiescent at initialization.}

In this work, to efficiently resolve the neck singularity at pinch-off, we adopt a two-stage AMR strategy based on the Basilisk implementation \citep{basilisk_plateau}. At early times, refinement (levels 7–10) is guided by wavelet-estimated errors in the interface, velocity, and strain-rate fields \citep{van_hooft_towards_2018}. Once the neck radius approaches the grid scale, refinement is restricted to the interface: cells are refined to level 15, then 
the minimun grid size $\Delta x= \mathcal{L}/2^{15}$,
while non-interface regions are derefined to level 10. This approach ensures both accuracy near the singularity and computational efficiency.


{Previous viscoplastic simulations in the Basilisk framework employed a viscosity-capped regularised Bingham model \citep{sanjay_bursting_2021}, in which the apparent viscosity is bounded by a prescribed maximum value to remove the singularity at vanishing strain rate. Here,}
the Herschel--Bulkley constitutive law is implemented via a regularised formulation in which the apparent viscosity is smoothed through an $\varepsilon$-dependent function to avoid singularities as $\|\boldsymbol{\mathcal{D}}\|\!\to\!0$, where $\boldsymbol{\mathcal{D}}$ denotes the rate-of-deformation tensor. Consequently, the yield surface cannot be sharply resolved, since $\|\boldsymbol{\mathcal{D}}\|$ remains strictly positive for any finite $\varepsilon>0$, although regions where $\|\boldsymbol{\mathcal{D}}\|$ falls below a prescribed threshold can be interpreted as approximately unyielded. In all simulations presented here, we set $\varepsilon=10^{-6}$; {the influence of $\varepsilon$ on the predicted regime maps is examined in Appendix~A}. 
{The implementation was validated at two levels and presented in Appendix~A. First, the constitutive discretization was verified against analytical solutions for steady planar Poiseuille flow of a Herschel–Bulkley fluid, confirming accurate recovery of yielded and plug regions as well as asymptotic consistency with respect to the regularization parameter. Second, a Newtonian thread-breakup benchmark   was performed, demonstrating that the solver recovers the expected Newtonian inertial-capillary dynamics in the limit 
($\mathcal{J}=0,\,n=1$), together with additional regularization-sensitivity checks.}

\section{Results}

\begin{figure*}
    \centering
    \includegraphics[width=\linewidth]{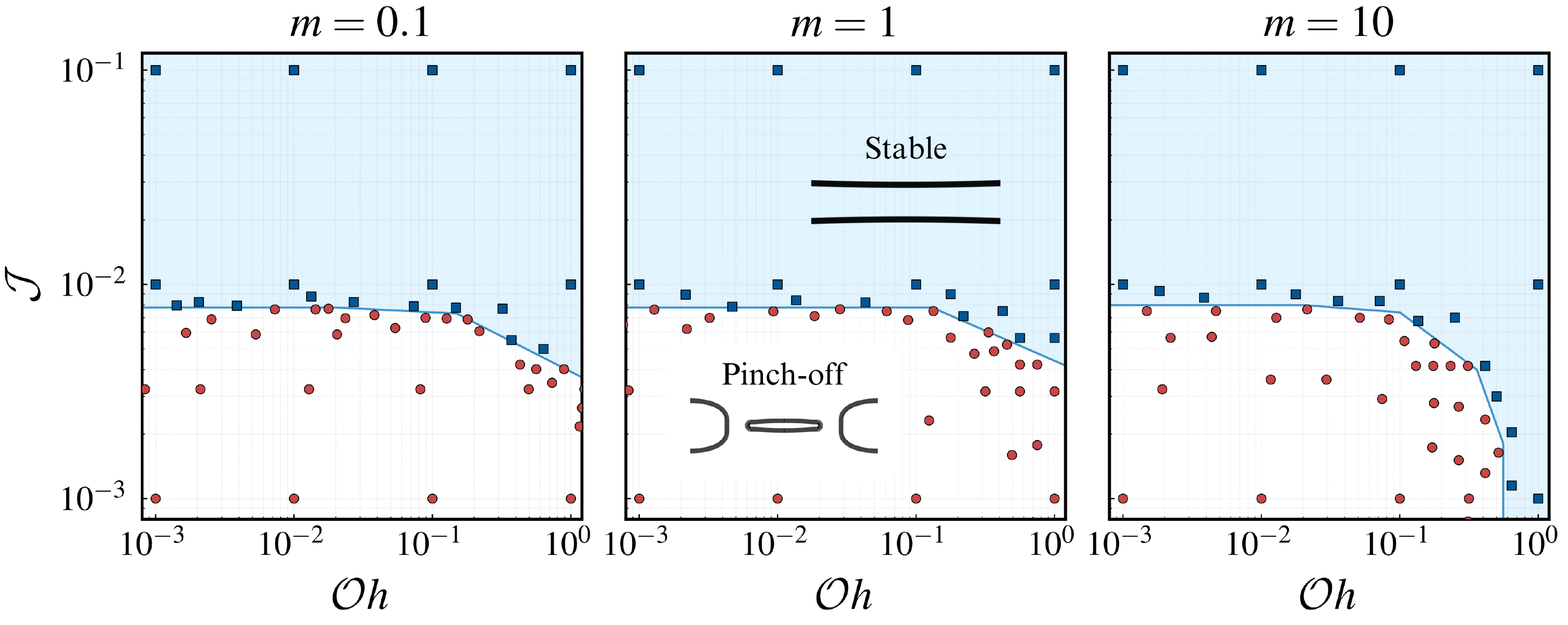}
    \vspace{0cm} 
    \hbox to \linewidth{ 
        \hspace{0.2\linewidth} (a)  
        \hspace{0.27\linewidth} (b)  
        \hspace{0.27\linewidth} (c)  
        \hfill 
    }
    \caption{Regime maps in the $\mathcal{J}$–$\mathcal{O}h$ plane for $m=(0.1,1,10)$, shown in panels (a)-(c). Red circles denote breakup, where the thread thins to pinch–off; blue squares denote stable threads that persist without rupture. Each dot corresponds to a simulation; solid blue lines are guides to the eye indicating the breakup boundary.}
    \label{fig:regime-maps}
\end{figure*}

Figure \ref{fig:regime-maps} shows a regime map in the $(\mathcal{O}h, \mathcal{J})$ space for different viscosity ratios, $m = 0.1, 1, 10$, with $n = 1$, $\rho_r = 1$, \CRCA{and $\delta=0.1$}.
The maps reveal the competing roles of capillarity, yield stress, and viscosity in thread thinning. At low $\mathcal{J}$, capillary forces dominate, and the thread thins continuously until breakup. Increasing $\mathcal{J}$ stabilizes the thread as the yield stress suppresses extensional flow. The critical yield number, $\mathcal{J}_c \approx 8\times10^{-3}$, below which pinch-off occurs, is largely independent of the viscosity ratio and of $\mathcal{O}h < 10^{-1}$, indicating that breakup is set primarily by the balance between capillary and yield stresses. For large Ohnesorge numbers ($\mathcal{O}h \gg 1$ for $m=1$ and $\mathcal{O}h \gg 5\times 10^{-1}$ for $m=10$), viscous forces dominate and, together with yield stress, suppress capillary-driven rupture. The regularization parameter $\varepsilon$ smooths the yield transition in the viscoplastic model; as $\varepsilon \to 0$, $\mathcal{J}_c$ converges to the ideal sharp yield stress criterion, \textcolor{black}{supporting the numerical robustness of the reported threshold for the prescribed initial condition.}
{All regime maps are constructed from simulations evolved up to $t=200$, corresponding to more than two orders of magnitude larger than the fastest capillary time scale. Cases classified as stable satisfy $h_{\min}(t) \ge 0.8\, R_0$ over the entire simulation window and exhibit no sustained monotonic decay of the minimum radius.}



\begin{figure*}
\centering
\includegraphics[width=\linewidth]{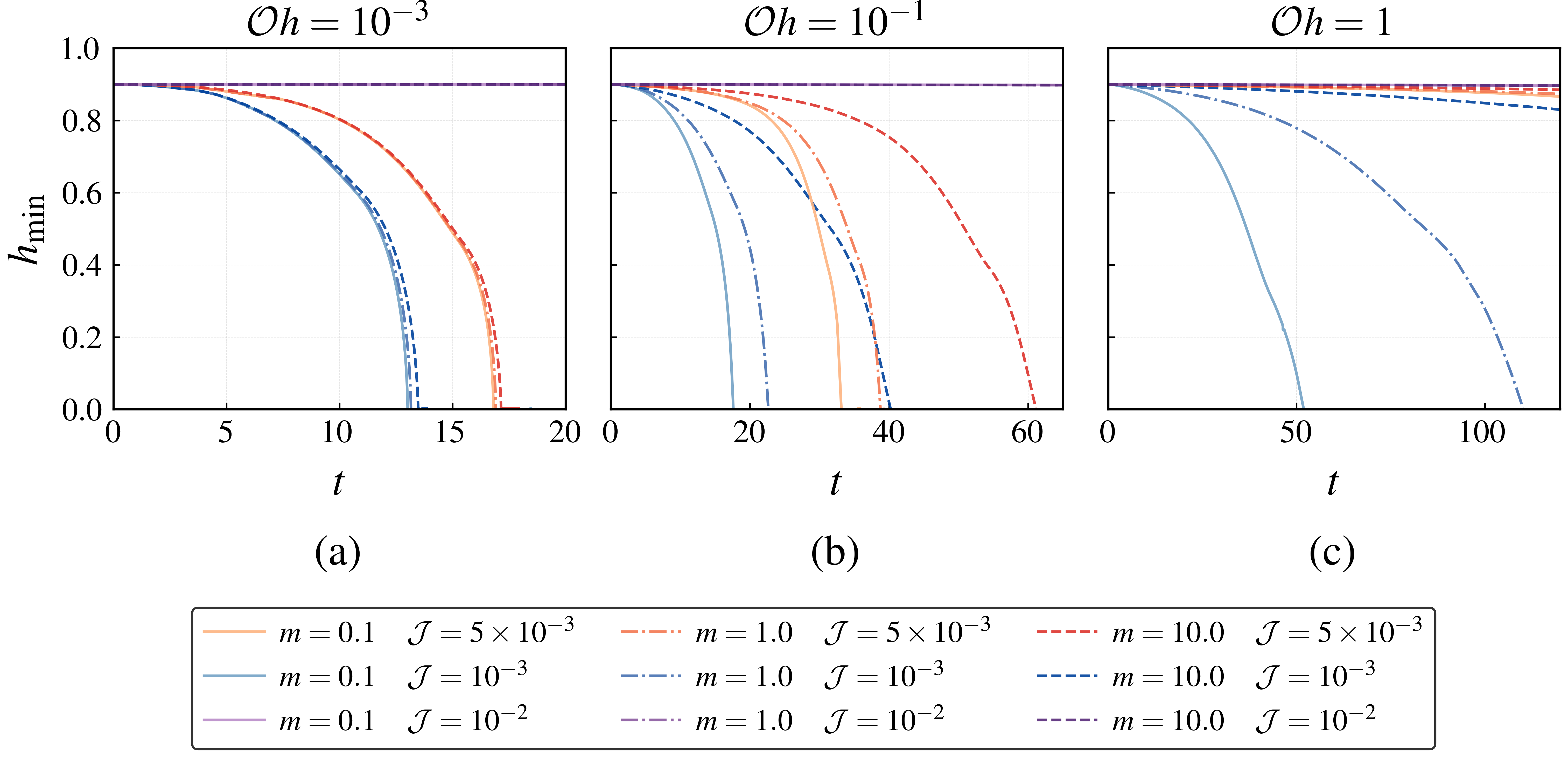}
\caption{Temporal evolution $h_{\min}$  over time up to $t=160$ for $\mathcal{O}h=(10^{-3},10^{-1},10^0)$ corresponding to panels (a)-(c), respectively.}
\label{fig:temp_evo}
\end{figure*}

While the regime map in figure \ref{fig:regime-maps} characterizes the final state of the thread, to elucidate the influence of  $m$ and $\mathcal{J}$ on the filament thinning dynamics, we examine the temporal evolution of the minimum thread radius $h_{\min}(t)$ across three Ohnesorge numbers: $\mathcal{O}h = (10^{-3},10^{-1}, 10^{0}$). For $\mathcal{O}h = 10^{-3}$, corresponding to the inertia-dominated regime in (nearly) inviscid threads, all cases with $\mathcal{J}<\mathcal{J}_c$ collapse onto a single curve independent of $m$, as viscous effects are negligible and the dynamics are governed by the inertial-capillary balance (refer to figure \ref{fig:temp_evo}a). When $\mathcal{J}>\mathcal{J}_c$ , the yield stress becomes comparable to the capillary stress and suppresses filament thinning, preventing breakup entirely. At $\mathcal{O}h = 10^{-1}$, viscous stresses begin to contribute appreciably to the force balance, and the thinning of the thread  exhibits a clear dependence on $m$ (shown in figure \ref{fig:temp_evo}b). Increasing the viscosity ratio  {progressively} delays pinch-off time, as the yield stress of the surrounding fluid resists interface motion. At $\mathcal{O}h = 1$, the dynamics become strongly viscosity-dominated, resulting in the slow down of the necking   across all cases (see figure \ref{fig:temp_evo}c). 
High $m$ and $\mathcal{J}$ combinations yield  states where $h_{\min}$ asymptotes to a finite value by $t = 160$. While we do not observe topological breakup within this time frame, these cases indicate that the system may asymptotically approach a non-breakup state.

\begin{figure}
    \centering
    \includegraphics[width=0.8\linewidth]{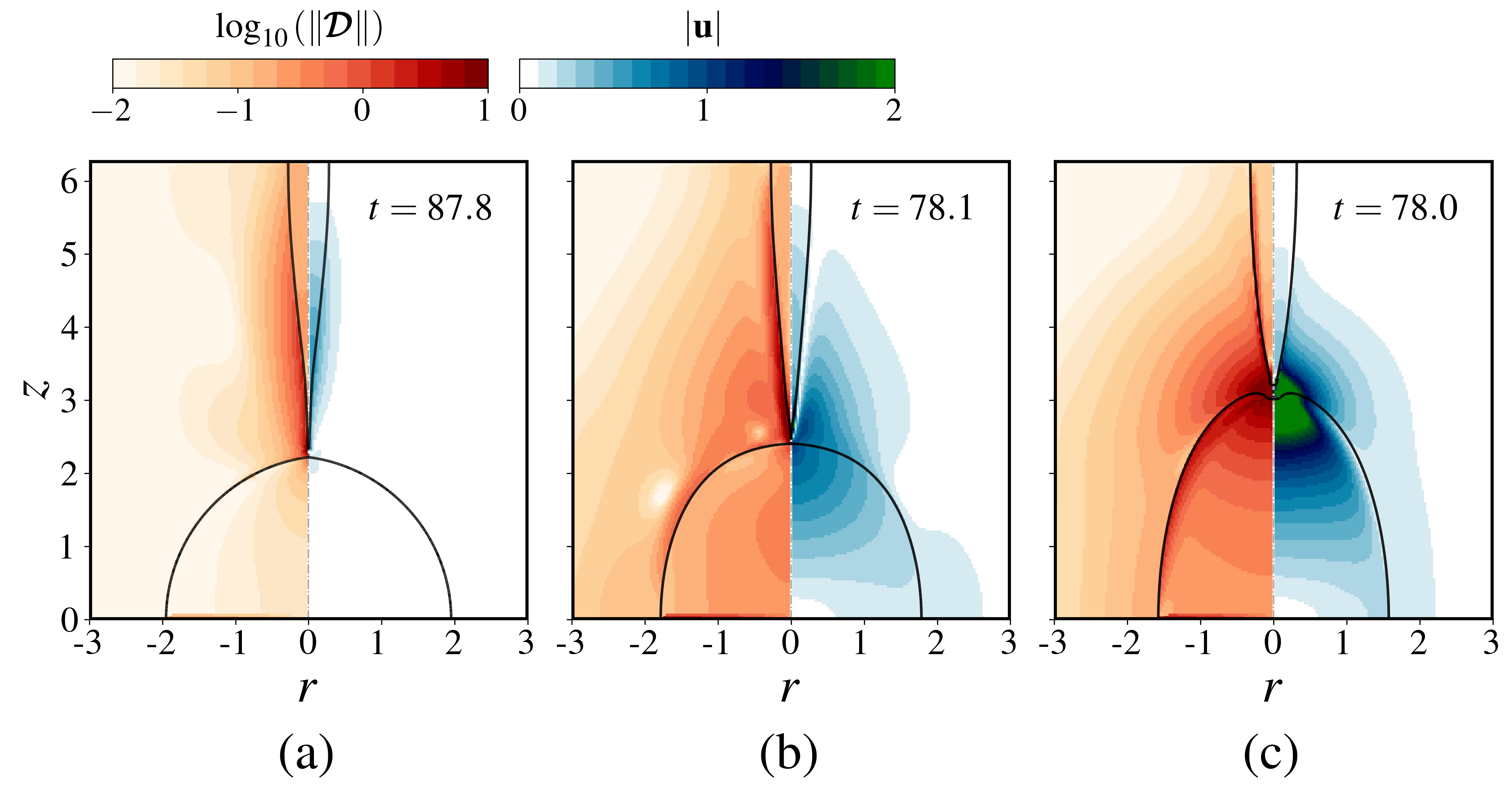}
    \caption{ Snapshots of the flow field {immediately after pinch-off ($t \gtrsim t_b$)}. Each panel shows the velocity field (right) and the deformation-rate magnitude $||\boldsymbol{\mathcal{D}}||$ (left) for:   (a) $\mathcal{O}h=1.204$, $\mathcal{J}=2.651\times10^{-3}$, $t_b=87.726$ (b) $\mathcal{O}h=2.812\times10^{-1}$, $\mathcal{J}=6.045\times10^{-3}$, $t_b=78.054$ (c) $\mathcal{O}h=7.210\times10^{-4}$, $\mathcal{J}=7.775\times10^{-3}$, $t_b=77.979$. Other parameters are identical to figure \ref{fig:regime-maps}.} 
    \label{fig:breakup_view_m0.1}
\end{figure}

To elucidate the thinning dynamics, figure \ref{fig:breakup_view_m0.1} presents the velocity field (right) and the deformation-rate magnitude $||\boldsymbol{\mathcal{D}}||$ (left), together with the interface position, for representative cases near the transition boundary at $m=0.1$ and various $\mathcal{O}h$. Snapshots are taken near the capillary singularity (e.g., the break-up time $t_b$ is reported in each panel), highlighting the flow structure.
In the high-$\mathcal{O}$h regime ($\mathcal{O}h = 1.204$, figure \ref{fig:breakup_view_m0.1}a), viscous stresses dominate, leading to significant resistance that suppresses fluid acceleration. This results in markedly reduced interface velocities and a delayed breakup, indicative of a viscously damped pinch-off.
In contrast, the low-$\mathcal{O}h$ regime ($\mathcal{O}h = 7.21 \times 10^{-4}$, figure \ref{fig:breakup_view_m0.1}c), where inertia and surface tension govern the dynamics, the interface accelerates rapidly without viscous damping, reaching high velocities at the neck, followed by a pronounced recoil after pinch-off—a hallmark of inertial-capillary-dominated breakup (see Supplemental information).
The intermediate case ($\mathcal{O}h = 2.812\times10^{-1}$, figure \ref{fig:breakup_view_m0.1}b) exhibits non-negligible viscous effects that partially suppress inertial motion. The interface velocities are moderate, and the breakup time is comparable to the low-$\mathcal{O}$h case, but without the post-breakup rebound, highlighting the transitional nature of the flow. 
%

We note that at high-$\mathcal{O}h$ values and large $m$, a secondary neck forms near the thread center, eventually leading to satellite droplets (see Supplemental Information). This occurs because the increased solid-like behavior of the external viscoplastic fluid restricts deformation near the thread ends, forcing flow and stress to localize in the center.
The complete temporal evolution for all these  cases is provided in the Supplemental Information.

Figure \ref{fig:flow-index-regime-map} shows the effect of the flow index $n$ on regime boundaries for $m=1$ and $\rho_r=1$. At low $\mathcal{O}h$, all curves collapse to $\mathcal{J}_c \approx 8\times 10^{-3}$, indicating that viscous effects are negligible and the rupture threshold is set solely by the balance of capillary and yield stresses, independent of $n$. As $\mathcal{O}h$ increases, viscous effects become comparable to capillarity, and the influence of $n$ emerges. In shear-thinning fluids ($n<1$), the local viscosity in the neck decreases under high strain rates, reducing viscous resistance, accelerating thinning, and shifting the breakup boundary downward and to the left of $n=1$. In shear-thickening fluids ($n>1$), the effective viscosity grows with strain rate, enhancing resistance to necking and moving the boundary upward relative to $n=1$. Together, these boundaries provide a unified framework for how viscoplastic medium can controls the thinning of Newtonian filaments.

\textcolor{black}{
To understand the stability boundary dependence on $\mathcal{O}h$, 
we consider a scaling balance between the capillary driving force and the stresses that resist thinning. The capillary forcing is set by the curvature difference along the thread, $\Delta \tilde{\kappa}$, since a uniform curvature would only produce a uniform Laplace pressure and would not drive axial motion. For $n=1$, this balance may be written as
\begin{equation}
\Delta \tilde\kappa \sim B(m)\mathcal{J} + C(m)\, \mathcal{O}h\, \dot{\tilde\epsilon}.
\end{equation}
here $\dot{\tilde \epsilon}$ is the dimensionless strain rate scaled by the inertio-capillary timescale, and $B(m)$ and $C(m)$ are constants. 
For the imposed sinusoidal initial condition, the curvature variation is controlled by the axial curvature of the interface. Since differentiating a cosine perturbation twice introduces a factor of $\tilde{k}^2$, the initial curvature difference scales as
$\Delta \tilde\kappa \approx \delta{k}^2.$ Physically, this means shorter wavelengths produce larger initial curvature gradients.
Thus, the stress balance becomes
\begin{equation}
   \delta \tilde{k}^2
   \sim   B(m) \mathcal{J} + C(m) Oh \dot{\tilde\epsilon}.
\end{equation}
In the regime where $\mathcal{O}h \dot{\tilde\epsilon} \ll 1$, the yield threshold is set by curvature alone:
\begin{equation}
\mathcal{J}_c \sim  B(m) \delta \tilde{k}^2.
\label{eq:critical_J_delta_k}
\end{equation}
When $\mathcal{O}h \dot{\tilde\epsilon} \gtrsim 1$, viscosity stabilizes the system; thus, reducing the yield stress needed as
\begin{equation}
\mathcal{J}_c \sim B(m) \delta\tilde{k}^2 - C(m) \mathcal{O}h \dot{\tilde\epsilon}.    
\end{equation}
Physically, low $\mathcal{O}h$ means the onset is controlled by the capillary-yield balance alone. At higher Oh, viscosity contributes to the stabilization, so the yield stress needed to stop rupture decreases with Oh (as observed in our numerical simulations).  Thus the onset of $Oh$-dependence at $\mathcal{O}h=O(10^{-1})$ is consistent with the small curvature forcing associated with the imposed perturbation.}

\begin{figure} \centering
\includegraphics[width=0.5\linewidth]{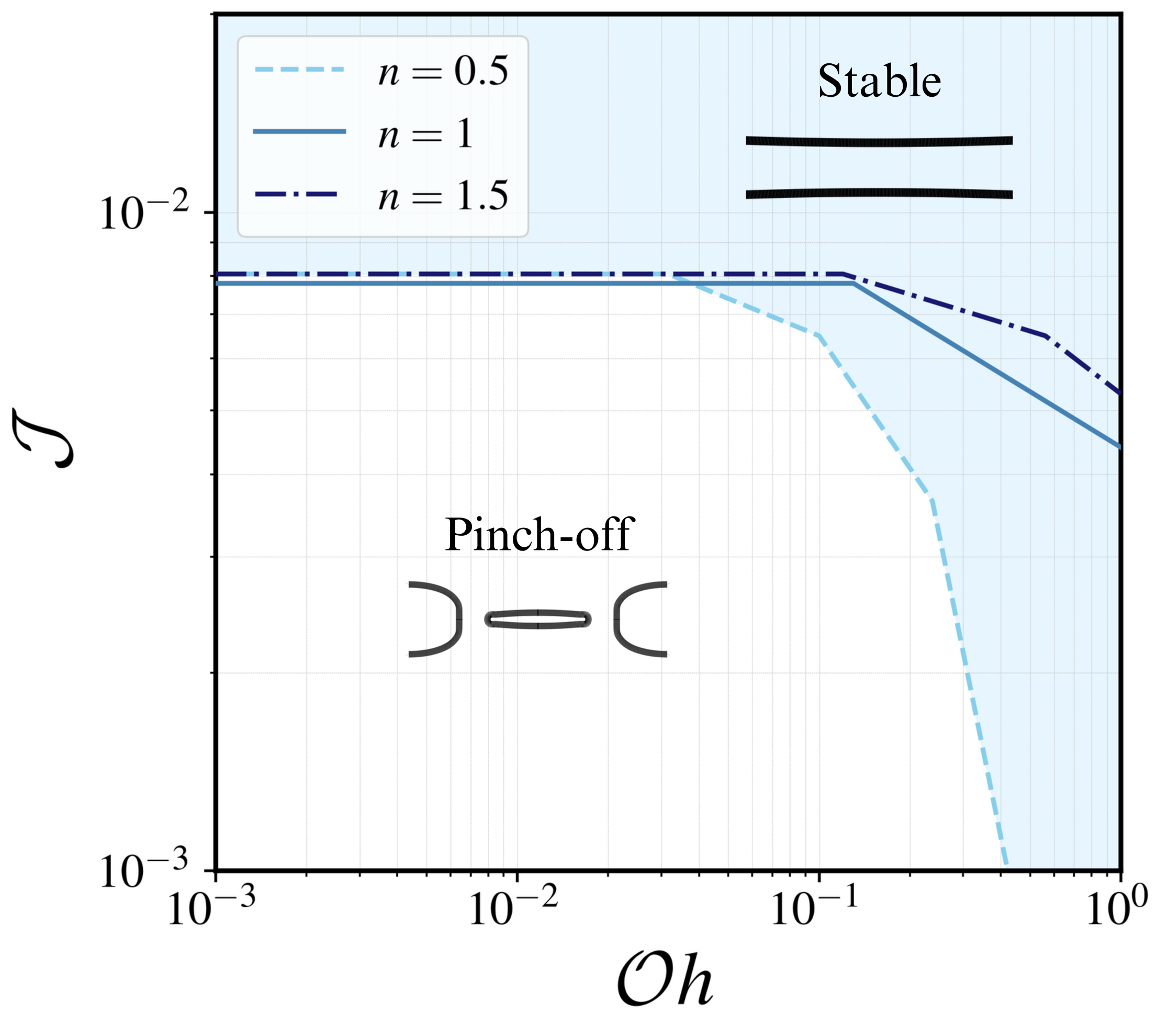}
\caption{Regime map boundaries for thread breakup in a viscoplastic fluid in the $(\mathcal{O}h, J)$ plane, for shear-thinning ($n=0.5$), Newtonian ($n=1$), and shear-thickening ($n=1.5$) cases with $m=1$. 
}
\label{fig:flow-index-regime-map}

\end{figure}

\begin{figure}
\centering
\begin{tabular}{cc}
\includegraphics[width=0.48\textwidth]{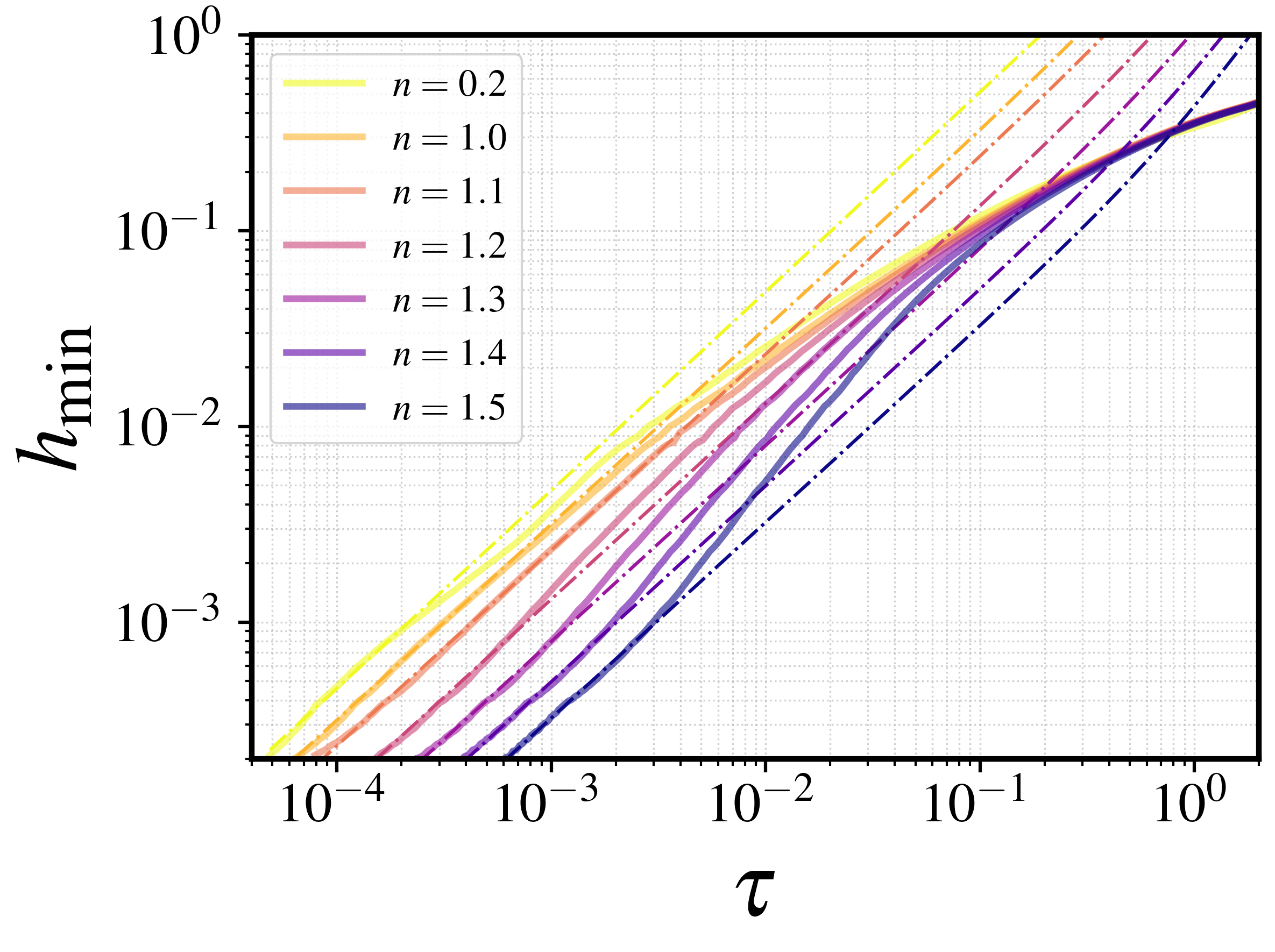} &
\includegraphics[width=0.48\textwidth]{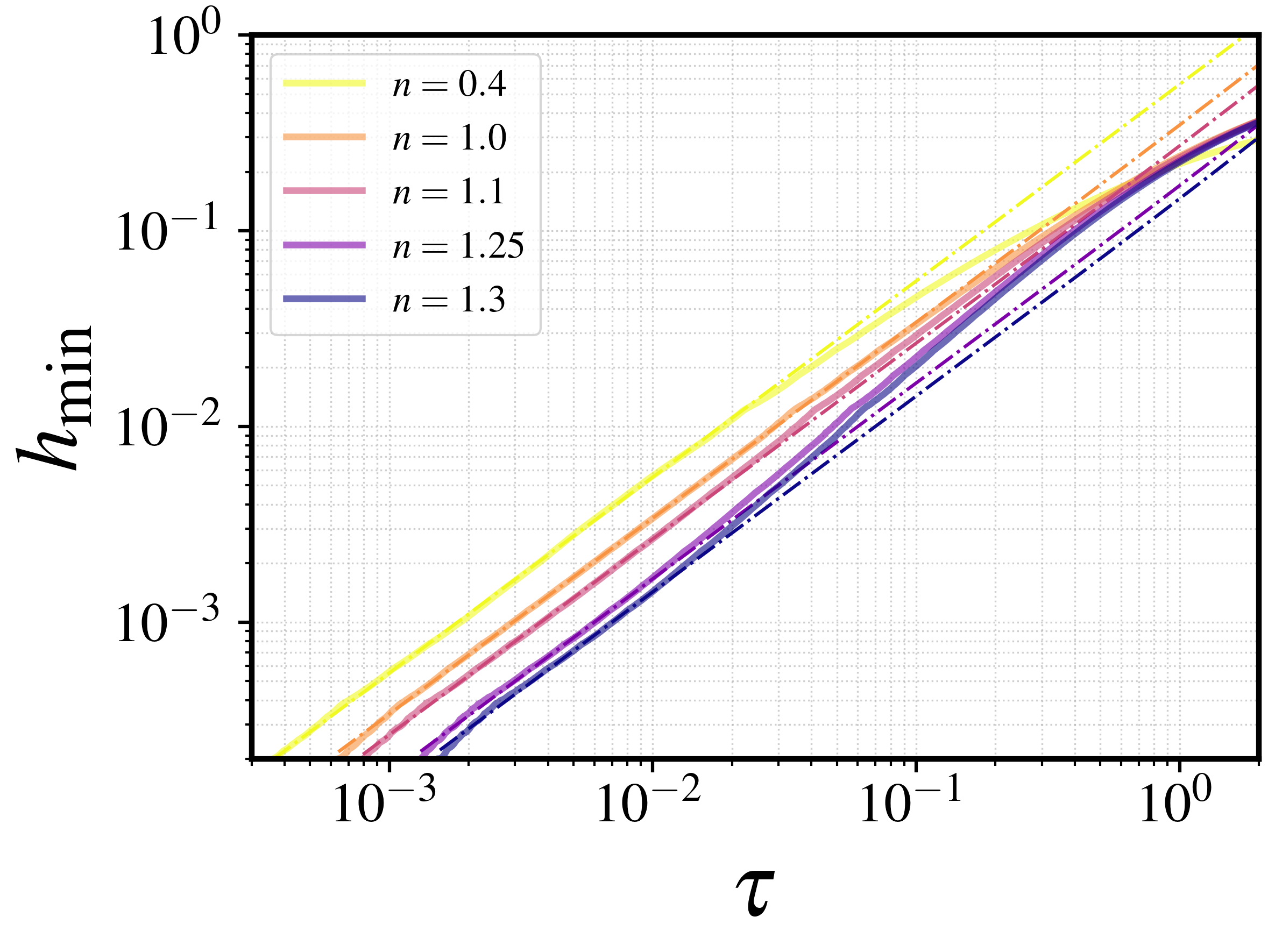}\\
(a) & (b) \\
\end{tabular}
\caption{Evolution of the minimum radius $h_{\min}$ as a function of $\tau$ for different $n$. Panel (a): $\mathcal{O}h=10^{-2}$, $\mathcal{J}=5.62\times 10^{-3}$; panel (b): $\mathcal{O}h=10^{-1}$, $\mathcal{J}=3.16\times 10^{-3}$. Solid line annotates simulation data while dashed line represents theoretical prediction according to the scaling law.}
\label{fig:scaling}
\end{figure}

\textcolor{black}{}

\subsection{Scalings for capillary breakup 
}

To understand the relative roles and interplay of surface tension, viscosity, and yield stress, \CRCA{we extend the Newtonian scaling arguments of \citet{eggers_universal_1993} and \citet{lister_capillary_1998} to the case in which the surrounding medium obeys a Herschel--Bulkley law.
We consider a slender axisymmetric thread in the $(z,r,t)$ coordinate system, with predominantly axial motion, $u=u(z,t)$, and interface shape $h=h(z,t)$. The reduced balances follow from a perturbation analysis of the constitutive equations; the complete derivation is given in Appendix~C. The resulting scalings are obtained by balancing the capillary pressure gradient against the Herschel--Bulkley resistance, which contains a rate-independent yield contribution
, $\mathcal{T}_y$,
and a rate-dependent viscous contribution.
$K_b\dot{\gamma}^{n}$.
These two contributions give rise to the terms proportional to $\mathcal{T}_y\tau/K_b$ and $\tau^{1-n}$ in the expressions below. Combining the corresponding velocity scale with the kinematic thinning relation over the remaining time to breakup, $\tau$, yields estimates for the maximum axial velocity, $u_{\max}$, the characteristic axial length scale, $L_c$, and the minimum thread radius, $h_{\min}$, for $n>0$:}

\begin{equation}
u_{\max} \sim \left(\frac{\sigma }{{K_b}}\right) \left[\left(\frac{\mathcal{T}_y \tau}{{K_b}}\right) + \tau^{1-n}\right]^{-1/2} \quad 
\label{eq:scaling_umax}
\end{equation}

\begin{equation}
L_{c} \sim 
\left(\frac{\sigma \tau }{{K_b}}\right)
\left(\frac{\mathcal{T}_y \tau}{{K_b}}\right)^{-\frac{1}{2}} + \left(\frac{\sigma \tau }{{K_b}}\right)  \tau^{\frac{n-1}{2}} 
\quad 
\label{eq:scaling_length}
\end{equation}

\begin{equation}
h_{\min} \sim \left(\frac{\sigma \tau}{\mu_{\mathrm{eff}}}\right) \left[\left(\frac{\mathcal{T}_y \tau}{{K_b}}\right) + \tau^{1-n}\right]
\quad 
\label{eq:scaling_hmin}
\end{equation}
\vspace{0.5em}  

It can be shown that a substitution of $n = 1$ recovers the predicted scalings of \citet{lister_capillary_1998} exactly. The main assumption underlying this analysis is that, during the late stages of thinning, the system reaches a regime in which capillary, viscous, and inertial forces are in balance, as demonstrated previously \citep{lister_capillary_1998}. 

Figure~\ref{fig:scaling} shows the evolution of the minimum thread radius $h_{\min}$ as a function of the time to breakup, $\tau = t_b - t$, for different flow indices $n$. Figure~\ref{fig:scaling}a corresponds to $\mathcal{O}h = 10^{-2}$, $\mathcal{J} = 5.62 \times 10^{-3}$, and figure~\ref{fig:scaling}b to $\mathcal{O}h = 10^{-1}$, $\mathcal{J} = 3.16 \times 10^{-3}$. The dashed lines indicate reference slopes of the theoretical predictions given by \textcolor{black}{equation~\ref{eq:scaling_hmin}}. Across the tested range $0.5 \leq n \leq 1.5$, the simulations collapse well onto the theoretical predictions, confirming that the thinning rate is governed by the balance between capillary pressure and the generalized shear-rate–dependent viscosity. For $n<1$ (shear-thinning), filaments thin more rapidly than in the Newtonian case ($n=1$), while for $n>1$ (shear-thickening), thinning is retarded, consistent with the shifts observed in the regime maps.

\subsection{\CRCA{Effect of thread geometry on $\mathcal{J}_c$, and comparison with experiments.}}
\label{sec: geom_Jc}

\CRCA{
In this section we analyze the sensitivity of the plastocapillary number to the  selected initial conditions. Specifically, we discuss the effect of the axial domain length and the  initial perturbation amplitude.}

\begin{figure}
\centering
\includegraphics[width=\textwidth]{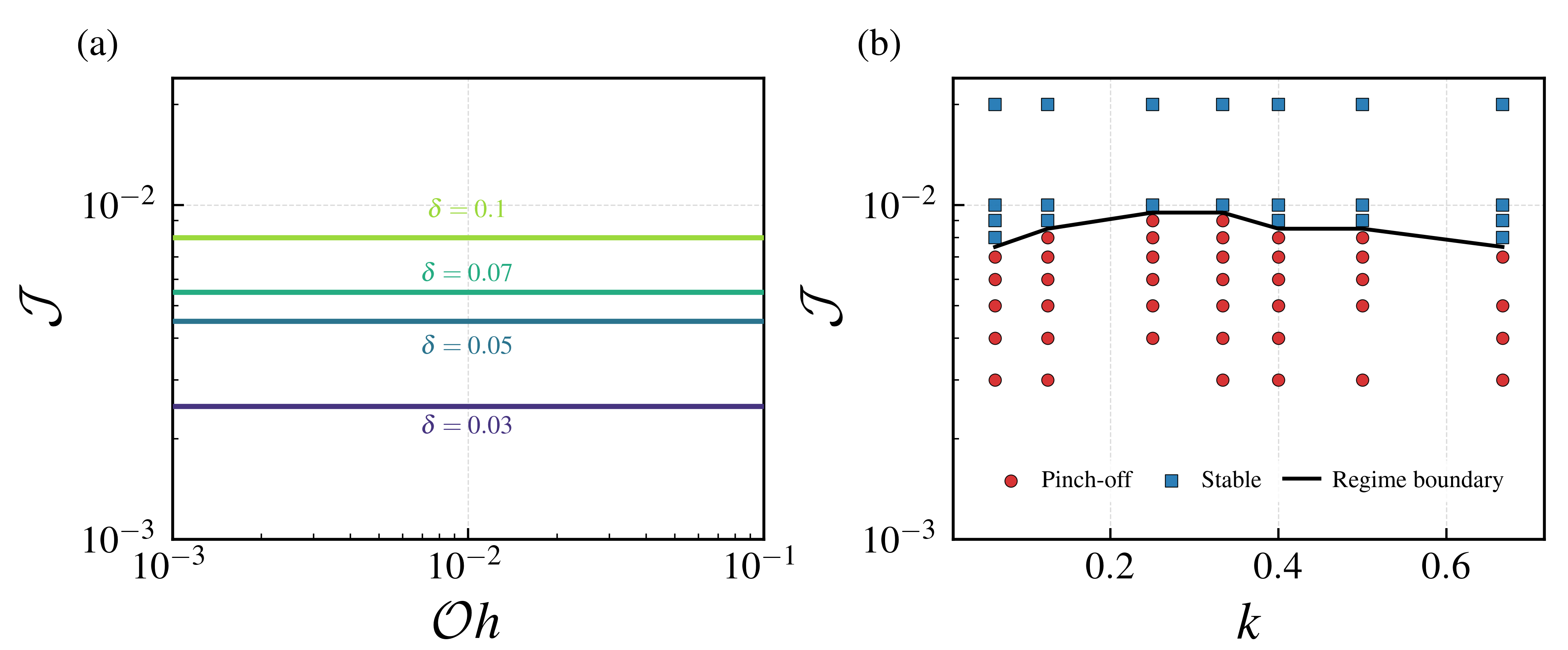}\\
\caption{
\textcolor{black}{
Sensitivity of the regime boundary to the imposed perturbation $\delta$ and wavenumber $k$, corresponding to panels (a) and (b), respectively.
For panel (b): $\mathcal{O}h=10^{-3}$, $\delta=0.1$, $m=1$, and $n=1$.}}
\label{fig:gemetry}
\end{figure}

\textcolor{black}{
Figure~\ref{fig:gemetry}a shows the influence of the initial perturbation amplitude $\delta$, we computed the regime boundary between arrest and pinch-off in the $\mathcal{J}$–$\mathcal{O}h$  plane  for different values of $\delta$ when $m=1$, $n=1$ and low $\mathcal{O}h$ . 
Decreasing $\delta$ shifts the  critical $\mathcal{J}_c$ for breakup to lower values.
Over the range considered, reducing $\delta$ from
0.1 to 0.03 lowers $\mathcal{J}_c$,  by approximately a factor of three, while the dependence on $\mathcal{O}h$ remains weak.
%
Yielding occurs when capillary stresses overcome the yield stress (as shown in equation~\ref{eq:critical_J_delta_k}). Since the maximum curvature induced by the initial perturbation scales with its amplitude,
the apparent critical yield--capillary ratio scales proportionally with $\delta$. Smaller perturbations therefore require a smaller $\mathcal{J}_c$ to initiate yielding, leading to the observed downward shift of the  threshold.
This behaviour reflects the finite-amplitude (activation-type) nature of the problem. Once thinning has started, the dynamics follows equations~\ref{eq:scaling_umax}--\ref{eq:scaling_hmin}.
}

\textcolor{black}{
Figure \ref{fig:gemetry}b shows the effect of the perturbation wavelength, or equivalently the ligament length, on whether the ligament breaks up or remains stable for
 $\mathcal{O}h=10^{-3}$, $m=1$,  $n=1$, and $\delta=0.1$.
 We consider the range ${1}/{16}<k<{1}/{1.5}$ ( ${3\pi}/{2}<{\lambda}/{2}<16\pi$).  (We note that the
rest of this study uses the base value $k=1/2$).
Over this wavenumber  range,
the dependence of 
the $\mathcal{J}_c$ on the ligament length  is weak and non-monotonic. In all
cases, the transition remains confined to the  interval
$0.007<\mathcal{J}_c<0.009$, indicating that the breakup threshold is relatively insensitive to the imposed wavenumber.
At small $k$, corresponding to longer 
wavelengths, the axial-curvature contribution to the initial disturbance is weaker, since $h_{zz}\sim \delta k^2$.
This is consistent with the slightly smaller values of $\mathcal{J}_c$ observed at the lowest wavenumbers.}

\textcolor{black}{
Finally, we compare the critical value of $\mathcal{J}_c$ obtained in the present simulations with the experimental transition reported by \citet{hossain_critical_2025} for long, large-aspect-ratio ligaments in a carbopol fluid.
In the experiments, the transition between breakup and non-breakup occurs at approximately $\mathcal{J}_c\simeq 0.2$, whereas in the present simulations we find a much smaller value, $ 0.0075 \lesssim \mathcal{J}_c \lesssim 0.0095 $ when varying the length of the ligament. Therefore, variations in the imposed wavelength, or domain length, at least over the range explored here, cannot account for the order-of-magnitude difference between the present simulations and the experimental critical value, $\mathcal{J}_c\simeq 0.2$. 
This leaves two likely explanations for the  discrepancy between their experiments and our  simulations. First, the present simulations are two-dimensional and axisymmetric, and may therefore not fully capture the three-dimensional disturbances present in the experiments. Such three-dimensional effects, particularly when combined with a possible larger perturbation amplitudes, could shift to larger   $\mathcal{J}_c$. Second, carbopol is not an ideal Herschel--Bulkley fluid,  it can exhibit viscoelastic effects that are not included in the present purely viscoplastic model. This suggests that modelling carbopol solely as a Herschel--Bulkley fluid may be insufficient to quantitatively reproduce the experiments. Incorporating elastic effects, for example through a Saramito--Herschel--Bulkley model, is therefore an important direction for future work.
}


\section{Conclusions}

We have shown that the pinch-off of a Newtonian thread in a viscoplastic medium is governed by a  rupture threshold, primarily set by the balance of capillary and yield stresses, largely independent of viscosity ratio at low Ohnesorge numbers.
Regime maps in the $(\mathcal{O}h, \mathcal{J})$ plane reveal distinct outcomes and delineate the transitions between arrested and pinch-off regimes. Similar qualitative behavior is observed for both shear-thinning and shear-thickening cases.
\textcolor{black}{We find that the critical plastocapillary number, $\mathcal{J}_c$, is not universal, but depends on the imposed initial condition. This dependence arises because the initial perturbation sets the curvature forcing that drives thinning, in which weaker initial curvature variations require a smaller yield stress to arrest the motion. The simulations also show that increasing $\mathcal{O}h$ lowers $\mathcal{J}_c$, since viscous stresses provide an additional stabilizing contribution. Thus, the transition between arrest and pinch-off is controlled not only by the material yield stress, but also by the initial thread shape and the relative importance of viscous stresses.}
After the onset of thinning, we extend the  framework of \citet{lister_capillary_1998} to account for thinning in a viscoplastic medium. The resulting scaling laws capture the evolution of the minimum thread radius and maximum axial velocity in close agreement with the simulations.

\textcolor{black}{
Finally, our work shows  
discrepancy on $\mathcal{J}_c$ with the experimental observations reported by \citet{hossain_critical_2025}.
The comparison  suggests that quantitative agreement with carbopol experiments may require accounting for effects absent from the present model, including larger or three-dimensional perturbations and the viscoelastic response of carbopol. Therefore, while the present simulations capture the capillary--yield-stress balance controlling arrest, a purely axisymmetric Herschel--Bulkley description is likely insufficient for direct quantitative comparison with the experiments. This is an interesting avenue that should be pursued. 
}

Beyond advancing fundamental understanding, this framework provides a basis for predicting and controlling embedded fluid threads in additive manufacturing and related technologies. The results also offer insight into broader capillary-driven interfacial flows, including pinhole formation in bursting bubbles \citep{vatsalsy_github_burstingbubble_vp,ghaemi2025burstingbubblesherschelbulkleyfluids}, droplet impact, and ligament retraction \citep{yang2026}, where the competition between thinning, arrest, and breakup plays a central role. \\


Declaration of Interests. The authors report no conflict of interest. \\

\subsection*{Acknowledgments}
This research used the Delta advanced computing and data resource which is supported by the National Science Foundation (award OAC 2005572) and the State of Illinois. CRCA acknowledges Randy Ewoldt for planting the seed of this idea. \CRCA{We gratefully acknowledge the three anonymous reviewers for their constructive comments and insightful suggestions, which have helped improve the clarity and quality of the manuscript.}

{\section*{Appendix A: Validation of Herschel--Bulkley model and parameter dependence}}
\label{append: A}


{To assess the accuracy of the $\varepsilon$-regularized viscoplastic formulation, we examine steady pressure-driven planar Poiseuille flow of a generalized Herschel--Bulkley fluid and compare the numerical results with the corresponding analytical profiles obtained from the classical one-dimensional reduction of the momentum balance for yield-stress fluids \citep{panaseti2018pressure}. The computations are performed in a two-dimensional domain of unit height with periodic boundary conditions in the streamwise direction. The wall-normal coordinate is defined so that $\tilde{y}=0$ denotes the solid wall, where no-slip is imposed, and $\tilde{y}=1$ denotes the symmetry plane at the channel centerline, where a free-slip symmetry condition is applied. 
All variables are nondimensionalized using the channel half-height $h$ and the wall shear stress $\tau_w$ as characteristic length and stress scales, with a reference viscosity $\mu$ to scale the shear rate, so that $\tilde{\mathcal T}=\mathcal T/\tau_w$, $\tilde{\dot\gamma}=\mu \dot\gamma/\tau_w$, and $\tilde K$ is the corresponding dimensionless consistency index, $\tilde K = K\,\tau_w^{\,n-1}/\mu^{\,n}$.}

{ A constant streamwise forcing, equivalent to a uniform pressure gradient, generates a fully developed unidirectional flow. The Stokes limit is considered and inertial effects are neglected. Under these conditions, the streamwise momentum equation reduces to
\begin{equation}
\frac{d \tilde{\mathcal{T}}}{d\tilde{y}} = -\,\frac{d\tilde{p}}{d\tilde{x}},
\end{equation}
which, for a constant nondimensional pressure gradient $d\tilde{p}/d\tilde{x} = 1$ together with zero shear at the centerline, gives the linear shear-stress distribution
\begin{equation}
\tilde{\mathcal{T}}(\tilde{y}) = 1 - \tilde{y}.
\end{equation}
The yielded region is defined by $|\tilde{\mathcal{T}}(\tilde{y})|>\tilde{\mathcal{T}}_y$, leading to a yield surface located at
\begin{equation}
\tilde{Y} = 1 - \tilde{\mathcal{T}}_y.
\end{equation}
For $0 \le \tilde{y} < \tilde{Y}$, the Herschel--Bulkley constitutive law
\begin{equation}
\tilde{\mathcal{T}} = \tilde{\mathcal{T}}_y + \tilde{K} \tilde{\dot{\gamma}}^{\,n}
\end{equation}
provides the shear-rate profile
\begin{equation}
\tilde{\dot{\gamma}}(\tilde{y})
=
\left(\frac{\tilde{\mathcal{T}}(\tilde{y})-\tilde{\mathcal{T}}_y}{\tilde{K}}\right)^{1/n}
=
\left(\frac{\tilde{Y}-\tilde{y}}{\tilde{K}}\right)^{1/n},
\end{equation}
while $\tilde{\dot{\gamma}}=0$ in the plug region $\tilde{Y} \le \tilde{y} \le 1$. The velocity follows from direct integration with the boundary condition $\tilde{u}(0)=0$,
\begin{equation}
\tilde{u}(\tilde{y})
=
\int_0^{\tilde{y}} \tilde{\dot{\gamma}}(\tilde{y})\, d\tilde{y}
=
\frac{n}{n+1}
\left(\frac{1}{\tilde{K}}\right)^{1/n}
\left[
\tilde{Y}^{\frac{n+1}{n}}
-
(\tilde{Y}-\tilde{y})^{\frac{n+1}{n}}
\right],
\qquad \tilde{y}<\tilde{Y},
\end{equation}
and is constant for $\tilde{y} \ge \tilde{Y}$.
In the Bingham limit ($n=1$), these relations reduce to
\begin{equation}
\tilde{\dot{\gamma}}(\tilde{y}) = \frac{\tilde{Y} - \tilde{y}}{\tilde{K}},
\qquad
\tilde{u}(\tilde{y}) = \frac{1}{\tilde{K}}
\left(
\tilde{Y}\tilde{y} - \frac{\tilde{y}^2}{2}
\right),
\qquad \tilde{y}<\tilde{Y},
\end{equation}
with a uniform plug velocity above the yield surface.
The Newtonian reference case, corresponding to $\tilde{\mathcal{T}}_y=0$, $n=1$, and $\tilde{K}=1$, is obtained from $\tilde{\mathcal{T}}=\tilde{\dot{\gamma}}$ and gives $\tilde{\dot{\gamma}}(\tilde{y})=1-\tilde{y}$ and $\tilde{u}(\tilde{y})=\tilde{y}-\frac{1}{2}\tilde{y}^2$.
}

\begin{figure}
\centering
\includegraphics[width=0.7\linewidth]{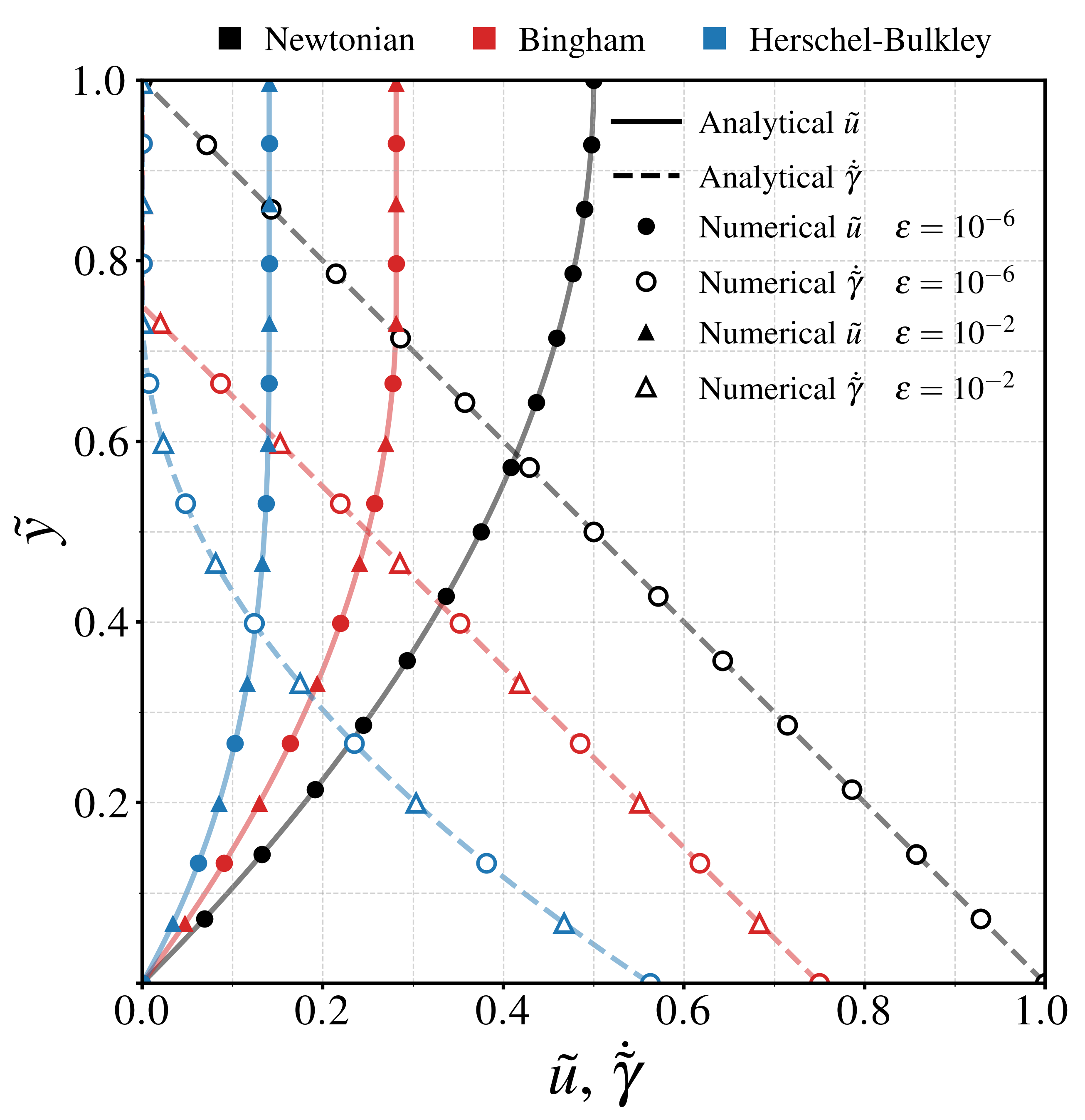}

\caption{{Validation of the $\varepsilon$-regularized formulation using steady pressure-driven planar Poiseuille flow.
Solid and dashed lines denote the analytical solutions of the velocity $\tilde{u}$ and shear rate $\dot{\tilde{\gamma}}$ profiles as functions of the wall-normal coordinate $\tilde{y}$,
while circular and triangular markers represent the numerical results obtained with $\varepsilon = 10^{-6}$ and $10^{-2}$, respectively.}}
\label{fig:HB_validation}
\end{figure}

{
Figure~\ref{fig:HB_validation} presents a direct comparison between the numerical solutions obtained using the $\varepsilon$-regularized constitutive law and the exact analytical solutions for steady, fully developed channel flow of Newtonian ($\tilde{\mu} = 1.0, \tilde{\mathcal{T}}_y=0$), Bingham $(\tilde{K} = 1.0, \tilde{\mathcal{T}}_y = 0.25, n=1.0)$, and Herschel--Bulkley $(\tilde{K} = 1.0, \tilde{\mathcal{T}}_y = 0.25, n=0.5)$ fluids. 
The comparison includes both the velocity profile $\tilde{u}(\tilde{y})$ and the shear-rate distribution $\tilde{\dot{\gamma}}(\tilde{y})$ across the  channel height, encompassing both yielded regions and the central plug where applicable. For the Newtonian case, the numerical solution recovers the exact parabolic velocity profile and linear shear-rate distribution dictated by the pressure–viscous balance. In the Bingham and Herschel–Bulkley cases, the computations reproduce the finite plug region, the analytically predicted plug width, and the piecewise structure of the solution. The shear rate vanishes within the unyielded core and exhibits a sharp transition at the yield surface, consistent with the theoretical stress distribution. We observe that the solutions are effectively independent of the regularization parameter, as the numerical results are indistinguishable for $\varepsilon \in [10^{-6},10^{-2}]$ when compared with the analytical predictions. These findings confirm that the regularized formulation used in this work consistently recovers the exact Newtonian limit and the analytical viscoplastic solutions in both yielded and unyielded regions.
}

\begin{figure}
\centering
\begin{tabular}{cc}
\includegraphics[width=0.45\linewidth]{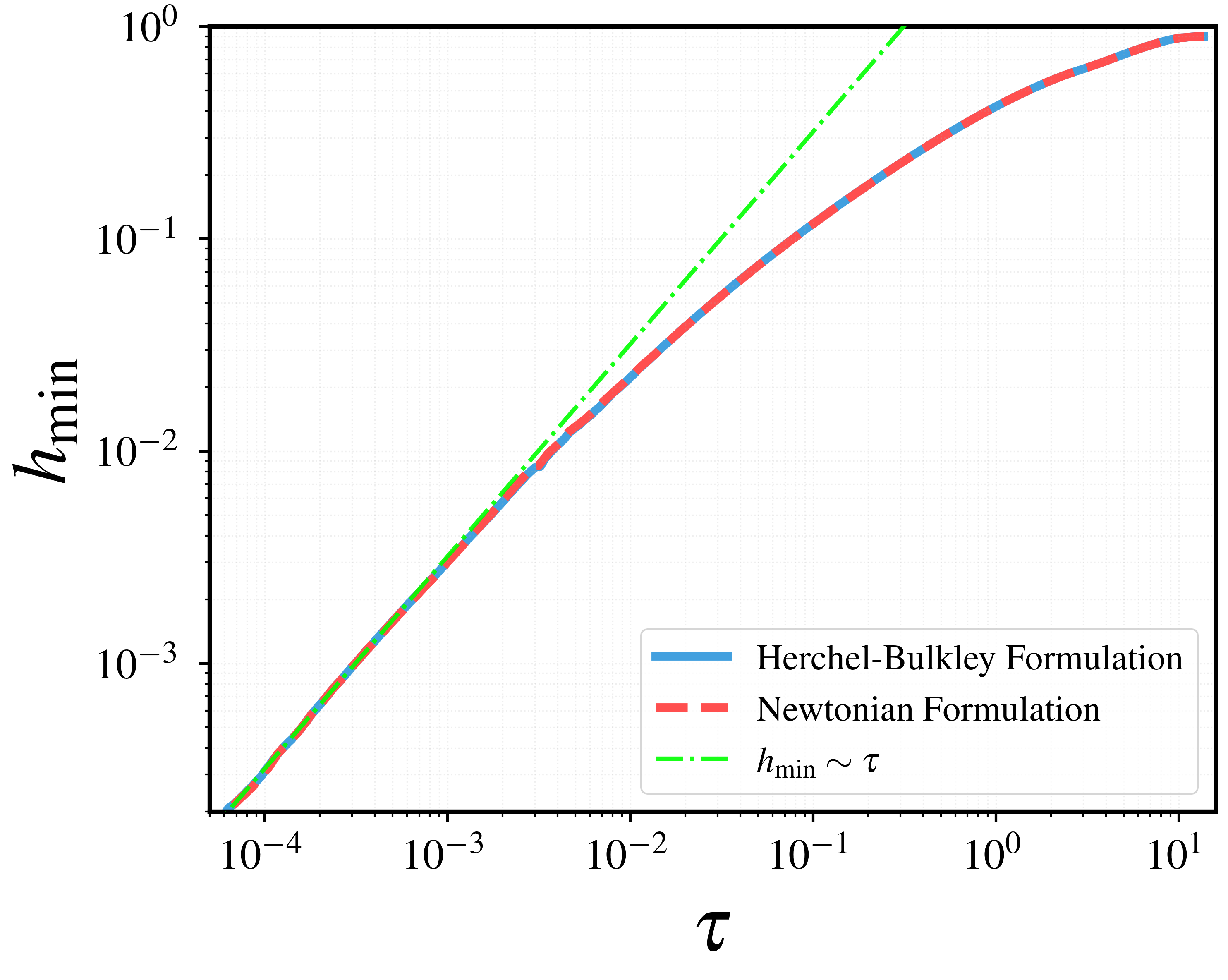}&
\includegraphics[width=0.45\linewidth]{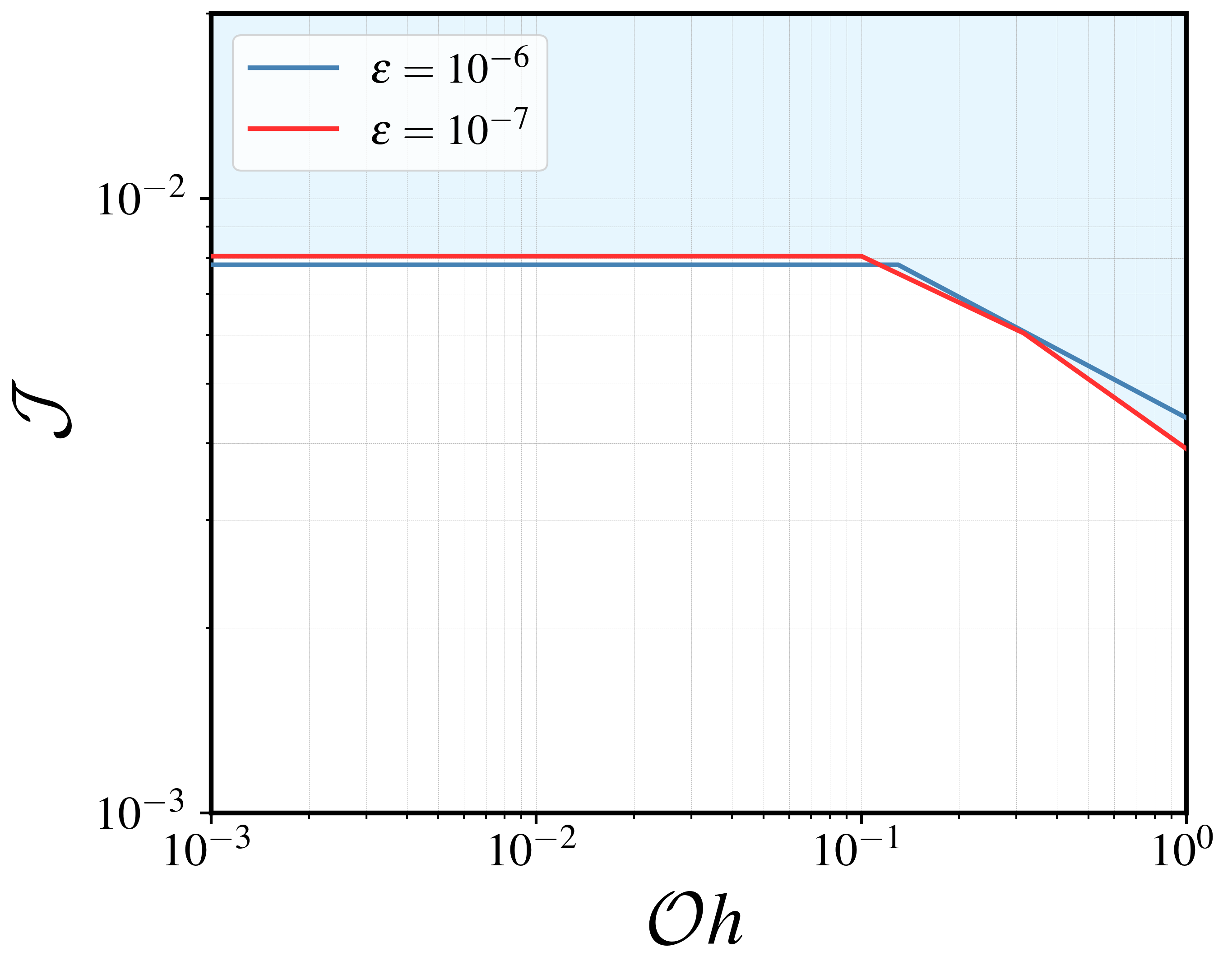}\\
 (a)    &  (b)\\
\end{tabular}
\caption{
{(a) 
Thread thinning: comparison between the classical Newtonian solution (red) and the $\varepsilon$-regularized formulation in the zero-yield limit (blue), demonstrating recovery of the Newtonian behavior at
 $\mathcal{O}h=10^{-2}$.  
(b) Regime maps in the $\mathcal{J}$–$\mathcal{O}h$ plane for $m = 1$ obtained using two regularization parameters, $\varepsilon = 10^{-6}$ and $\varepsilon = 10^{-7}$}.
}
\label{fig_apped}
\end{figure}

{To further validate the Herschel--Bulkley implementation, we simulate the breakup of a Newtonian thread characterized by $\mathcal{O}h = 10^{-2}$, surrounded by a Newtonian fluid. Supplemental figure~\ref{fig_apped}a shows the temporal evolution of the neck radius for both the Newtonian case and its Herschel--Bulkley counterpart with $\mathcal{J} = 0$, $n=1$ and $m = 1$. The excellent agreement between both cases, including the recovery of the expected Newtonian scaling laws, confirms the accuracy of the Herschel--Bulkley formulation employed in this study. Supplemental figure~\ref{fig_apped}b illustrates the effect of the regularization parameter $\varepsilon$ for a limiting case near a regime boundary.
To assess the sensitivity of the results to the regularization used in the yield-stress model, we performed a systematic comparison using two regularization parameters, $\varepsilon = 10^{-6}$ and $\varepsilon = 10^{-7}$ (see figure~\ref{fig_apped}b). We find that both the stability boundaries in the $\mathcal{J}$-$\mathcal{O}h$ regime maps and the observed flow dynamics are insensitive to further reductions of $\varepsilon$ beyond $\varepsilon = 10^{-6}$.
}\\

\section*{\textcolor{black}{Appendix B: Validation  with \citet{beris_creeping_1985}. }}

\textcolor{black}{
We have validated  against the classical benchmark of \citet{beris_creeping_1985} for steady creeping flow past a sphere in a Bingham fluid. For this case, the solver was reduced to a single-phase axisymmetric configuration. The reference sphere radius, velocity and viscosity were set to $R_0=V_0=\mu_0=1$.
The computation was performed in a square axisymmetric domain of side length $L=16$, with the symmetry axis located at $y=0$
The sphere was represented using a
embedded-boundary formulation \citep{TAVARES}.
 Specifically, the solid boundary was defined by the level-set function
$\phi_s(x,y)=x^2+y^2-R_0^2$,
so that $\phi_s=0$ corresponds to the sphere surface and $\phi_s<0$ denotes the solid region.
}

\textcolor{black}{
The calculation was performed in a sphere-fixed frame. Thus, a uniform far-field velocity $u_x=-V_0$ was imposed, corresponding to inflow at the right boundary and outflow at the left boundary.
The pressure reference was set by imposing ($p=0$) at the left boundary.
At the upper boundary, the far-field streamwise velocity was imposed together with zero transverse velocity, while the lower boundary was treated as the axis of symmetry, with $u_y=0$ and $\partial_y u_x=0$. No-slip and no-penetration conditions were imposed on the embedded sphere, so that the velocity vanished on the solid surface in the sphere-fixed frame.
Finally, the same regularized constitutive form used in the main simulations of this work is retained, with $\epsilon=1\times10^{-5}$, and  $n=1$  to recover the Bingham fluid.
A quadtree mesh was used, with refinement levels $6\leq \ell \leq 11$ , and the region $x^2+y^2<(2.5R_0)^2$ surrounding the sphere was initially refined.
}

\textcolor{black}{
We compare the  drag coefficient $C_s$ and the corresponding $\mathcal{J}_c$, 
directly with the values reported in Table 2 of \citet{beris_creeping_1985}.
The drag coefficient is calculated as $C_s={F}/({6\pi\mu_0 V_0 R_0})$, where $F$ is calculated at the sphere as
 $     F =
  \left|
  \int_{S_s}
  \left(
  -p\mathcal{\boldsymbol{I}}
  +2\mu_{\mathrm{eff}}\mathcal{\boldsymbol{D}}
  \right)\boldsymbol{n}\cdot\boldsymbol{e}_x\,\mathrm{d}S
  \right|,$
here $S_s$ denotes the sphere surface, $\boldsymbol{n}$ is the outward normal, and the axisymmetric surface element is $\mathrm{d}S=2\pi r_b,\mathrm{d}l$.
Then, we calculate
$\mathcal{J}_c$ as $\mathcal{J}c={N_B}/{6C_s}$, where $N_B$ is the Bingham number used by \citet{beris_creeping_1985}.}

\begin{figure}
\centering
\includegraphics[width=0.98\linewidth]{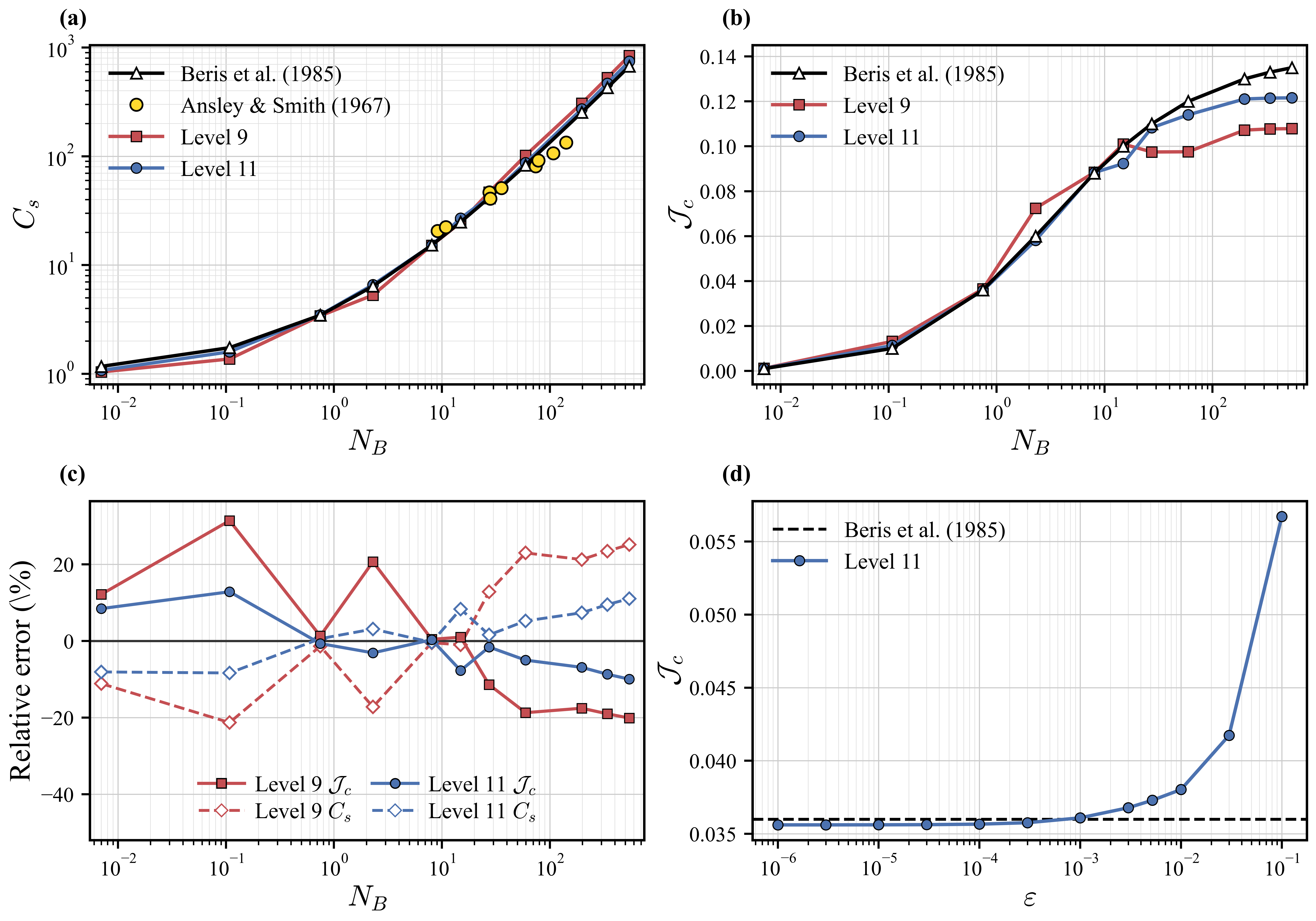}
\caption{
\textcolor{black}{
Comparison of the present Basilisk simulations with the  values reported in Table 2 of \citet{beris_creeping_1985}.
(a) Drag coefficient $C_s$ as a function of $N_B$ for refinement levels 9 and 11.
(b) Plastocapillary number $\mathcal{J}_c$ as a function of the Bingham number $N_B$ for refinement levels 9 and 11.
(c)  Relative errors in $\mathcal{J}_c$ and $C_s$
(d) Sensitivity of $\mathcal{J}_c$ to the regularization parameter $\varepsilon$ for $N_B=0.747$, with the dashed line denoting the  value reported by \citet{beris_creeping_1985}.}
}
\label{fig:beris_validation_combined}
\end{figure}

\textcolor{black}{
Table \ref{tab:beris_validation} and figure \ref{fig:beris_validation_combined} 
compare the present Basilisk simulations
with the results of  \citet{beris_creeping_1985}.   The  simulations reproduce the monotonic increase of the $C_s$ with Bingham number and show good overall agreement with both the numerical results of \citet{beris_creeping_1985} and the experimental measurements of \citet{Ansley}  (see figure \ref{fig:beris_validation_combined}a and \ref{fig:beris_validation_combined}b).
Increasing the maximum refinement level from 9 to 11 improves the agreement for both $C_s$ and $\mathcal{J}_c$ (see figures \ref{fig:beris_validation_combined}a-c). This suggests that the remaining discrepancies being associated with
the resolution of velocity gradients and stresses near the embedded boundary, which directly enter the force calculation. Further refinement would likely reduce these errors, but was not pursued because of the associated computational cost.}

\textcolor{black}{
We also tested the sensitivity to the regularization parameter, $\epsilon$, for $N_B=0.747$. For $\epsilon \lesssim 10^{-3}$, both $C_s$ and $\mathcal{J}_c$ 
become essentially insensitive to further decreases in
$\epsilon$ and approach values close to those reported by
\citet{beris_creeping_1985} (see panel \ref{fig:beris_validation_combined}d).
Larger values of $\epsilon$ lead to systematic deviations, consistent with over-regularization of the material response, particularly in regions of large stress gradients near the solid boundary.}

\textcolor{black}{
Finally, some discrepancy is expected because Basilisk represents the sphere through an embedded-boundary reconstruction on a Cartesian grid, whereas \citet{beris_creeping_1985} used a finite-element method. This difference affects both the geometric representation of the sphere and the numerical evaluation of stresses on its surface, and therefore directly influences the computed drag force and the inferred value of $\mathcal{J}_c$.
}

\begin{table}
\centering
\caption{
\textcolor{black}{Comparison between the our Basilisk simulations with level 11 and the values reported in Table 2 of \citet{beris_creeping_1985}.}}
\label{tab:beris_validation}
\begin{tabular}{rcc c ccc}
\toprule
& \multicolumn{2}{c}{Beris et al. (1985)} &
\multicolumn{2}{c}{Present simulations} &
\multicolumn{2}{c}{Relative difference} \\
\cmidrule(lr){2-3}\cmidrule(lr){4-5}\cmidrule(lr){6-7}
$N_B$ & $C_s$ & $\mathcal{J}_c$ & $C_s$ & $\mathcal{J}_c$ &
$\Delta C_s$ (\%) & $\Delta \mathcal{J}_c$ (\%) \\
\midrule
0.007 & 1.17  & 0.001 & 1.08   & 0.0011 &  -8.06 &   8.46 \\
0.747 & 3.46  & 0.036 & 3.48   & 0.0358 &   0.61 &  -0.66 \\
2.299 & 6.39  & 0.060 & 6.59   & 0.0581 &   3.12 &  -3.09 \\
8.047 & 15.24 & 0.088 & 15.19  & 0.0883 &  -0.34 &   0.35 \\
14.91 & 24.85 & 0.100 & 26.92  & 0.0923 &   8.34 &  -7.70 \\
27.36 & 41.45 & 0.110 & 42.12  & 0.1083 &   1.61 &  -1.57 \\
59.59 & 82.77 & 0.120 & 87.11  & 0.1140 &   5.24 &  -4.98 \\
197.5 & 253.2 & 0.130 & 271.89 & 0.1211 &   7.38 &  -6.87 \\
340.7 & 426.9 & 0.133 & 467.50 & 0.1215 &   9.51 &  -8.68 \\
544.6 & 672.3 & 0.135 & 746.58 & 0.1216 &  11.05 &  -9.94 \\
\bottomrule
\end{tabular}
\end{table}

\section*{Appendix C: Regular perturbation \& scaling analysis }

\subsection*{C1. 3D axisymmetric thinning of a free-surface flow
}

The viscous motion of an axisymmetric fluid column with a free surface, in the asymptotic limit where $u = u(z,t)$ and $h = h(z,t)$ was first analyzed by \citet{eggers_universal_1993}. The effect of the viscosity of the surrounding fluid was neglected and was subsequently analyzed by \citet{lister_capillary_1998}. This work considers the suspending medium as viscoplastic, as opposed to prior works, which have either neglected this contribution altogether or have considered it as Newtonian. The general constitutive equations describing this system are as follows.

Continuity:
\begin{equation}
    \frac{\partial u_{r}}{\partial r} + \frac{\partial u_{z}}{\partial z} + \frac{u_{r}}{r} = 0 
\end{equation}
$z$-component of the Navier-Stokes:
\begin{equation}
    \rho\left\lbrack \frac{\partial u_{z}}{\partial t} + u_{r}\frac{\partial u_{z}}{\partial r} + u_{z}\frac{\partial u_{z}}{\partial z} \right\rbrack = - \frac{\partial p}{\partial z} + \mu\left\lbrack \frac{\partial^{2}u_{z}}{\partial r^{2}} + \frac{\partial^{2}u_{z}}{\partial z^{2}} + \frac{1}{r}\frac{\partial u_{z}}{\partial r} \right\rbrack 
\end{equation}
$r$-component of the Navier-Stokes:
\begin{equation}
    \rho\left\lbrack \frac{\partial u_{r}}{\partial t} + u_{r}\frac{\partial u_{r}}{\partial r} + u_{z}\frac{\partial u_{r}}{\partial z} \right\rbrack = - \frac{\partial p}{\partial r} + \mu\left\lbrack \frac{\partial^{2}u_{r}}{\partial r^{2}} + \frac{\partial^{2}u_{r}}{\partial z^{2}} + \frac{1}{r}\frac{\partial u_{r}}{\partial r} - \frac{u_{r}}{r^{2}} \right\rbrack 
\end{equation}
Locus of the fluid neck:
\begin{equation}
    \frac{\partial h}{\partial t} + u_{z}\frac{\partial h}{\partial z} = \left. \ u_{r} \right|_{r = h} 
\end{equation}

\textcolor{black}{Here, $\rho$ and $\mu$ denote the density and Newtonian viscosity, $u_z$ is the velocity along the axis, $u_r$ is the velocity in the radial direction, and $p$ is the pressure. The boundary conditions are
\begin{equation}
    \mathbf{n}\cdot\boldsymbol{{T}}\cdot\mathbf{n}
    =
    -\sigma\left(\frac{1}{R_1}+\frac{1}{R_2}\right),
\end{equation}
and
\begin{equation}
    \mathbf{n}\cdot\boldsymbol{{T}}\cdot\mathbf{t} =   0.
\end{equation}
Here $\boldsymbol{{T}}$ is the stress tensor, $\mathbf{n}$ the outward
normal,  $\mathbf{t}$ the unit tangent vector, and $R_1$ and $R_2$ are the principal radii of curvature.
}





The assumed forms of the velocity and pressure profiles are as follows (only incorporate lowest-order radial disturbances). The forms of these equations show that the main transport occurs along the azimuthal and temporal coordinates, with the radial coordinate only contributing to disturbances.

\begin{equation}
    u_{z}(z,r,t) = u_{0}(z,t) + u_{2}(z,t)r^{2} + \ldots 
\end{equation}

\begin{equation}
    u_{r}(z,r,t) = - \frac{\partial u_{0}(z,t)}{\partial z} \cdot \frac{r}{2} - \frac{\partial u_{2}(z,t)}{\partial z} \cdot \frac{r^{3}}{4} - \ldots 
\end{equation}

\begin{equation}
    p(z,r,t) = p_{0}(z,t) + p_{2}(z,t)r^{2} + \ldots 
\end{equation}

\textcolor{black}{In the regular perturbation expansion, we retain the velocity components $u_z$ and $u_r$. For the one-dimensional slender-neck equations below, we write $u(z,t)\equiv u_0(z,t)$ for the leading-order axial velocity.}
\textcolor{black}{
The one-dimensional equations for the axial velocity $u(z,t)$ and neck radius $h(z,t)$ reads,
\begin{equation}
\rho\left(\frac{\partial u}{\partial t}  + u \frac{\partial u}{\partial z} \right)
=
-\sigma \frac{\partial \kappa}{\partial z}
-\rho g
+
3\mu \frac{1}{h^2} \frac{\partial}{\partial z} \left(h^2 \frac{\partial u}{\partial z} \right) ,
\label{eq:slender_momentum}
\end{equation}
together with the kinematic condition
\begin{equation}
\frac{\partial h}{\partial t} + u \frac{\partial h}{\partial z} 
=
-\frac{h}{2}\frac{\partial u}{\partial z}  .
\label{eq:slender_kinematic}
\end{equation}
Here, $\kappa=R_1^{-1}+R_2^{-1}$ denotes the interfacial curvature.}

\subsection*{C2. Thinning in Newtonian viscous medium} 

\noindent \textcolor{black}{
Here we consider the case of a constant external viscosity, 
described in \citet{lister_capillary_1998}. In this regime, before molecular dimensions are reached and while the fluid cylinder remains in the near-field limit where the lubrication approximation is valid, the dominant balance is between the axial gradient of the capillary pressure and the viscous resistance from both the draining liquid and the surrounding fluid. Inertial contributions are neglected in this near-field Stokes-flow limit. 
}
\textcolor{black}{In this case, axial capillary-pressure gradient should balance the internal viscous resistance in $z$ and external viscous resistance in $r$.
Thus,
\begin{equation}
\sigma \frac{\partial}{\partial z}\left(\frac{1}{h}\right)
\sim
\mu \frac{\partial^2 u}{\partial z^2}
\sim
\mu_{\mathrm{ext}} \frac{\partial^2 u}{\partial r^2}.
\end{equation}
We use the following dimensionless variables 
\begin{equation}
h=\tilde hH,\qquad
r=\tilde rR,\qquad
z=\tilde z L_c,\qquad
t=\tilde t\tau,\qquad
u=\tilde u u_{\max}=\tilde u\left(\frac{L_c}{\tau}\right).
\label{Nondimensionalization_Lister}
\end{equation}
Here, $H$ and $R$ are local neck and radial length scales used in the near-field scaling.
Note that the natural length and time scales derived in the Eggers regime no longer apply directly. Substituting the above scalings gives
\begin{equation}
\sigma
\frac{\partial}{\partial (\tilde zL_c)}
\left(\frac{1}{\tilde h H}\right)
\sim
\mu
\frac{\partial^2(\tilde u u_{\max})}{\partial (\tilde z L_c)^2}
\sim
\mu_{\mathrm{ext}}
\frac{\partial^2(\tilde u u_{\max})}{\partial (\tilde r R)^2}.
\end{equation}
Therefore,
\begin{equation}
\left(\frac{\sigma}{L_cH}\right)
\frac{\partial}{\partial \tilde z}\left(\tilde h^{-1}\right)
\sim
\left(\frac{\mu u_{\max}}{L_c^2}\right)
\frac{\partial^2 \tilde u}{\partial \tilde z^{2}}
\sim
\left(\frac{\mu_{\mathrm{ext}}u_{\max}}{R^2}\right)
\frac{\partial^2 \tilde u}{\partial \tilde r^{2}}.
\end{equation}
Using $u_{\max}\sim L_c/\tau$, this becomes
\begin{equation}
\left(\frac{\sigma}{L_cH}\right)
\frac{\partial}{\partial \tilde z}\left(\tilde h^{-1}\right)
\sim
\left(\frac{\mu}{L_c\tau}\right)
\frac{\partial^2 \tilde u}{\partial \tilde z^{2}}
\sim
\left(\frac{\mu_{\mathrm{ext}}L_c}{R^2\tau}\right)
\frac{\partial^2 \tilde  u}{\partial \tilde r^{2}}.
\end{equation}
Keeping all terms dimensionless requires
\begin{equation}
\frac{\sigma}{L_cH}
\sim
\frac{\mu}{L_c\tau},
\end{equation}
which gives
\begin{equation}
H\sim \frac{\sigma \tau}{\mu}.
\end{equation}
Equating the second and third prefactors gives
\begin{equation}
\frac{\mu}{L_c\tau}
\sim
\frac{\mu_{\mathrm{ext}}L_c}{R^2\tau},
\end{equation}
and therefore
\begin{equation}
R^2 \sim L_c^2\left(\frac{\mu_{\mathrm{ext}}}{\mu}\right)
\sim L_c^2 m,
\qquad
R\sim L_c m^{1/2},
\qquad
m=\frac{\mu_{\mathrm{ext}}}{\mu}.
\end{equation}
In the lubrication limit, $R\sim H$. Thus,
\begin{equation}
R\sim L_c m^{1/2}
\sim H
\sim \frac{\sigma \tau}{\mu},
\end{equation}
which implies
\begin{equation}
L_c \sim \frac{\sigma \tau}{\mu m^{1/2}}.
\end{equation}
Lastly, using the velocity scaling,
\begin{equation}
u_{\max}\sim \frac{L_c}{\tau},
\end{equation}
we obtain
\begin{equation}
u_{\max}\sim \frac{\sigma}{\mu m^{1/2}}.
\end{equation}
Hence, the near-field Stokes-flow scalings are
\begin{equation}
z\sim L_c\sim \frac{\sigma \tau}{\mu m^{1/2}},
\qquad
h\sim H\sim \frac{\sigma \tau}{\mu},
\qquad
u\sim u_{\max}\sim \frac{\sigma}{\mu m^{1/2}}.
\label{lister_Eq}
\end{equation}
These scalings agree with the work by \citet{lister_capillary_1998}.
}

\subsection*{C3. Thinning in viscoplastic  medium}

\SHU{%
We extend \citet{lister_capillary_1998} to our case
study.  The scaling is derived for the most general case, with the coordinate system of choice being $(z,r,t)$. It is assumed that $u=u(z,t)$ and $h=h(z,t)$. In the viscoplastic extension, we identify the local neck scale $H$ with the minimum thread radius, $H \sim h_{\min}$.
The dimensionless terms are identical to \ref{Nondimensionalization_Lister}.}


\textcolor{black}{
Beginning with the derivation of the effective viscosity factor
$m_{\mathrm{eff}}=\mu_{\mathrm{eff}}/K_b$, we write the Herschel--Bulkley
constitutive law in tensorial form as
\begin{equation}
    \mathcal{T}_{ij}
    =
    2\mu_{\mathrm{eff}}\mathcal{D}_{ij},
    \qquad
    \mu_{\mathrm{eff}}
    =
    \frac{\mathcal{T}_y}{\dot{\gamma}_{\mathrm{eff}}}
    +
    K_b\dot{\gamma}_{\mathrm{eff}}^{\,n-1},
    \qquad n>0,
    \label{eq:hb_law_appendix}
\end{equation}
where $\dot{\gamma}_{\mathrm{eff}}$ is the scalar effective shear rate. We define
\begin{equation}
    \dot{\gamma}_{\mathrm{eff}}
    =
    2\|\boldsymbol{\mathcal{D}}\|
    =   \sqrt{2\boldsymbol{\mathcal{D}}:\boldsymbol{\mathcal{D}}},
    \qquad
    \|\boldsymbol{\mathcal{D}}\|
    =
    \sqrt{
    \frac{1}{2}
    \boldsymbol{\mathcal{D}}:\boldsymbol{\mathcal{D}}
    }.
    \label{eq:effective_shear_rate_appendix}
\end{equation}
Thus, the effective viscosity may also be written as
\begin{equation}
    \mu_{\mathrm{eff}}
    =
    \frac{\mathcal{T}_y}{2\|\boldsymbol{\mathcal{D}}\|}
    +
    K_b
    \left(
    2\|\boldsymbol{\mathcal{D}}\|
    \right)^{n-1}.
    \label{eq:mu_eff_appendix}
\end{equation}
Dividing by $K_b$, we obtain
\begin{equation}
    m_{\mathrm{eff}}
    =
    \frac{\mu_{\mathrm{eff}}}{K_b}
    =
    \frac{\mathcal{T}_y}
    {2K_b\|\boldsymbol{\mathcal{D}}\|}
    +
    \left(
    2\|\boldsymbol{\mathcal{D}}\|
    \right)^{n-1},
    \qquad n>0.
    \label{eq:m_eff_general_appendix}
\end{equation}
For the one-dimensional slender-filament approximation, the axial velocity is
taken as $u_z=u(z,t)$. The corresponding radial velocity follows from
incompressibility and is approximately
\begin{equation}
    u_r
    =
    -\frac{r}{2}
    \frac{\partial u_z}{\partial z}.
    \label{eq:radial_velocity_appendix}
\end{equation}
Therefore, the leading-order components of the rate-of-deformation tensor are
\begin{equation}
    \mathcal{D}_{zz}
    =
    \frac{\partial u_z}{\partial z},
    \qquad
    \mathcal{D}_{rr}
    =
    -\frac{1}{2}
    \frac{\partial u_z}{\partial z},
    \qquad
    \mathcal{D}_{\theta\theta}
    =
    -\frac{1}{2}
    \frac{\partial u_z}{\partial z},
    \qquad
    \mathcal{D}_{rz}\simeq 0.
    \label{eq:D_components_appendix}
\end{equation}
The last approximation follows from the leading-order slender-filament
assumption, for which the dominant deformation is axial extension accompanied
by radial compression. Hence,
\begin{equation}
    \boldsymbol{\mathcal{D}}:\boldsymbol{\mathcal{D}}
    =
    \mathcal{D}_{zz}^{2}
    +
    \mathcal{D}_{rr}^{2}
    +
    \mathcal{D}_{\theta\theta}^{2}
    +
    2\mathcal{D}_{rz}^{2}
    \simeq
    \frac{3}{2}
    \left(
    \frac{\partial u_z}{\partial z}
    \right)^2.
    \label{eq:D_contract_appendix}
\end{equation}
Using the definition of the tensor norm,
\begin{equation}
    \|\boldsymbol{\mathcal{D}}\|
    =
    \sqrt{
    \frac{1}{2}
    \boldsymbol{\mathcal{D}}:\boldsymbol{\mathcal{D}}
    }
    =
    \frac{\sqrt{3}}{2}
    \left|
    \frac{\partial u_z}{\partial z}
    \right|.
    \label{eq:D_norm_slender_appendix}
\end{equation}
Therefore,
\begin{equation}
    \dot{\gamma}_{\mathrm{eff}}
    =
    2\|\boldsymbol{\mathcal{D}}\|
    =
    \sqrt{3}
    \left|
    \frac{\partial u_z}{\partial z}
    \right|.
    \label{eq:gamma_eff_slender_appendix}
\end{equation}
Substituting this result into the expression for $m_{\mathrm{eff}}$ gives
\begin{equation}
    m_{\mathrm{eff}}
    =
    \frac{\mathcal{T}_y}
    {K_b\sqrt{3}
    \left|
    \partial u_z/\partial z
    \right|}
    +
    \left(
    \sqrt{3}
    \left|
    \frac{\partial u_z}{\partial z}
    \right|
    \right)^{n-1},
    \qquad n>0.
    \label{eq:m_eff_slender_appendix}
\end{equation}
The velocity derivative may be nondimensionalized as follows:
\begin{equation}
    \frac{\partial u_z}{\partial z}
    =
    \frac{\partial(\tilde u_z u_{\max})}{\partial(\tilde z L_c)}
    =
    \left(\frac{u_{\max}}{L_c}\right)
    \frac{\partial \tilde u_z}{\partial \tilde z}
    =
    \left(\frac{L_c/\tau}{L_c}\right)
    \frac{\partial \tilde u_z}{\partial \tilde z}
    =
    \frac{1}{\tau}
    \frac{\partial \tilde u_z}{\partial \tilde z}.
\end{equation}
Therefore,
\begin{equation}
    \left|
    \frac{\partial u_z}{\partial z}
    \right|
    =
    \frac{1}{\tau}
    \left|
    \frac{\partial \tilde u_z}{\partial \tilde z}
    \right|.
\end{equation}
Using
\begin{equation}
    \dot{\gamma}_{\mathrm{eff}}
    =
    \sqrt{3}
    \left|
    \frac{\partial u_z}{\partial z}
    \right|,
\end{equation}
the effective viscosity factor becomes
\begin{equation}
    m_{\mathrm{eff}}
    =
    \frac{\mu_{\mathrm{eff}}}{K_b}
    =
    \frac{\mathcal{T}_y}{\sqrt{3}K_b}
    \left|
        \frac{1}{\tau}
        \frac{\partial \tilde u_z}{\partial \tilde z}
    \right|^{-1}
    +
    \left(\sqrt{3}\right)^{n-1}
    \left|
        \frac{1}{\tau}
        \frac{\partial \tilde u_z}{\partial \tilde z}
    \right|^{n-1},
    \qquad n>0.
\end{equation}
Equivalently,
\begin{equation}
    m_{\mathrm{eff}}
    =
    \frac{1}{\sqrt{3}}
    \left(
        \frac{\mathcal{T}_y\tau}{K_b}
    \right)
    \left|
        \frac{\partial \tilde u_z}{\partial \tilde z}
    \right|^{-1}
    +
    \left(\sqrt{3}\right)^{n-1}
    \tau^{1-n}
    \left|
        \frac{\partial \tilde u_z}{\partial \tilde z}
    \right|^{n-1},
    \qquad n>0.
\end{equation}
Therefore,
\begin{equation}
    m_{\mathrm{eff}}^{-1/2}
    =
    \left[
    \frac{1}{\sqrt{3}}
    \left(
        \frac{\mathcal{T}_y\tau}{K_b}
    \right)
    \left|
        \frac{\partial \tilde u_z}{\partial \tilde z}
    \right|^{-1}
    +
    \left(\sqrt{3}\right)^{n-1}
    \tau^{1-n}
    \left|
        \frac{\partial \tilde u_z}{\partial \tilde z}
    \right|^{n-1}
    \right]^{-1/2},
    \qquad n>0.
    \label{m_eq}
\end{equation}
If the regime of \citet{lister_capillary_1998} is considered, equations~\eqref{lister_Eq} are modified using \eqref{m_eq} as follows.
\\
Maximum velocity $(u_{\max})$.
Using \eqref{m_eq}, the maximum velocity scale becomes
\begin{equation}
    u_{\max}
    \sim
    \frac{\sigma}{K_b}
    m_{\mathrm{eff}}^{-1/2}
\end{equation}
Therefore,
\begin{equation}
    u_{\max}
    \sim
    \frac{\sigma}{K_b}
    \left[
    \frac{1}{\sqrt{3}}
    \left(
        \frac{\mathcal{T}_y\tau}{K_b}
    \right)
    \left|
        \frac{\partial \tilde u_z}{\partial \tilde z}
    \right|^{-1}
    +
    \left(\sqrt{3}\right)^{n-1}
    \tau^{1-n}
    \left|
        \frac{\partial \tilde u_z}{\partial \tilde z}
    \right|^{n-1}
    \right]^{-1/2},
    \qquad n>0.
\end{equation}
If the dimensionless velocity gradient is assumed to be order one,
$\left|\partial \tilde u/\partial \tilde z\right|=O(1)$, then, up to
order-one prefactors,
\begin{equation}
    u_{\max}
    \sim
    \frac{\sigma}{K_b}
    \left[
    \left(
        \frac{\mathcal{T}_y\tau}{K_b}
    \right)
    +
    \tau^{1-n}
    \right]^{-1/2},
    \qquad n>0.
\end{equation}
Maximum length scale $(L_c)$. Since $L_c\sim u_{\max}\tau$, we obtain
\begin{equation}
    L_c
    \sim
    \frac{\sigma \tau}{K_b}
    m_{\mathrm{eff}}^{-1/2}.
\end{equation}
Thus,
\begin{equation}
    L_c
    \sim
    \frac{\sigma \tau}{K_b}
    \left[
    \frac{1}{\sqrt{3}}
    \left(
        \frac{\mathcal{T}_y\tau}{K_b}
    \right)
    \left|
        \frac{\partial \tilde u_z}{\partial \tilde z}
    \right|^{-1}
    +
    \left(\sqrt{3}\right)^{n-1}
    \tau^{1-n}
    \left|
        \frac{\partial \tilde u_z}{\partial \tilde z}
    \right|^{n-1}
    \right]^{-1/2},
    \qquad n>0.
\end{equation}
If $\left|\partial \tilde u_z/\partial \tilde z\right|=O(1)$, this reduces to
\begin{equation}
    L_c
    \sim
    \frac{\sigma \tau}{K_b}
    \left[
    \left(
        \frac{\mathcal{T}_y\tau}{K_b}
    \right)
    +
    \tau^{1-n}
    \right]^{-1/2},
    \qquad n>0.
\end{equation}
The corresponding limiting balances are
\begin{equation}
    L_c
    \sim
    \frac{\sigma \tau}{K_b}
    \left(
        \frac{\mathcal{T}_y\tau}{K_b}
    \right)^{-1/2},
    \qquad
    \frac{\mathcal{T}_y\tau}{K_b}
    \gg
    \tau^{1-n},
\end{equation}
and
\begin{equation}
    L_c
    \sim
    \frac{\sigma \tau}{K_b}
    \tau^{(n-1)/2},
    \qquad
    \tau^{1-n}
    \gg
    \frac{\mathcal{T}_y\tau}{K_b}.
\end{equation}
Minimum neck radius scale $(H)$. If the neck scale is modified through the same effective-viscosity factor as
\begin{equation}
    H
    \sim
    \frac{\sigma \tau}{\mu_{\mathrm{ext}}}
    m_{\mathrm{eff}},
\end{equation}
then
\begin{equation}
    H
    \sim
    \frac{\sigma \tau}{\mu_{\mathrm{ext}}}
    \left[
    \frac{1}{\sqrt{3}}
    \left(
        \frac{\mathcal{T}_y\tau}{K_b}
    \right)
    \left|
        \frac{\partial \tilde u_z}{\partial \tilde z}
    \right|^{-1}
    +
    \left(\sqrt{3}\right)^{n-1}
    \tau^{1-n}
    \left|
        \frac{\partial \tilde u_z}{\partial \tilde z}
    \right|^{n-1}
    \right],
    \qquad n>0.
\end{equation}
If $\left|\partial \tilde u_z/\partial \tilde z\right|=O(1)$, then
\begin{equation}
    H
    \sim
    \frac{\sigma \tau}{\mu_{\mathrm{ext}}}
    \left[
    \left(
        \frac{\mathcal{T}_y\tau}{K_b}
    \right)
    +
    \tau^{1-n}
    \right],
    \qquad n>0.
\end{equation}
}

Using the previous results, we obtain the equations 3.5-3.7.
It may be shown that for $n = 1$, these scaling laws identically collapse to those of \citet{lister_capillary_1998}.

{\section*{Appendix \textcolor{black}{D}: Nomenclature}}

{We present a summary of the main symbols and definitions used throughout this work. Table~\ref{tab:nomenclature}  lists  variables and parameters, including the dimensionless groups governing the dynamics.
}

\begin{table}
\caption{Nomenclature used throughout the manuscript. The table lists the principal variables, material parameters and
Subscripts $l$ and $b$ denote the Newtonian liquid thread and the surrounding medium, respectively.}
\label{tab:nomenclature}
{\footnotesize
\begin{multicols}{2}
\noindent
\begin{tabular}{@{}l p{0.32\textwidth}@{}}
\toprule
\textbf{Symbol} & \textbf{Definition} \\
\midrule
$\boldsymbol{\mathcal{D}}$ & Rate-of-deformation tensor\\
$f$ & Volume fraction (VOF variable) \\
$\boldsymbol{f}_\sigma$ & Surface tension force \\
$h$ & Initial interface radius profile \\
$h_{\min}$ & Minimum interface radius \\
\textcolor{black}{$k$} & \textcolor{black}{Wavenumber} \\
$\mathcal{J}$ & Plastocapillary number, $\mathcal{T}_y R_0 / \sigma$ \\
$K_b$ & Medium consistency index \\
$\mathcal{L}$ & Dimensionless domain length \\
$L_0$ & Dimensional domain length \\
$L_c$ & Characteristic length scale \\
$m$ & Normalized viscosity ratio \\
$m_{\mathrm{eff}}$ & Medium effective viscosity ratio \\
$n$ & Medium power-law index \\
$\mathbf{n}$ & Unit normal vector to the interface \\
\textcolor{black}{$\mathbf{t}$} & \textcolor{black}{Unit tangent vector to the interface}\\
$\mathcal{O}h$ & Ohnesorge number, $\mu_l / \sqrt{\rho_l \sigma R_0}$ \\
$p$ & Pressure \\
$r,\,z$ & Radial and axial coordinate \\
$R_0$ & Initial interface radius \\
\bottomrule
\end{tabular}

\columnbreak 

\noindent
\begin{tabular}{@{}l p{0.32\textwidth}@{}}
\toprule
\textbf{Symbol} & \textbf{Definition} \\
\midrule
$t$ & Time \\
$t_b$ & Breakup time \\
$t_i$ & Inertial-capillary time scale \\
$\mathcal{T}_y$ & Medium yield stress \\
$\boldsymbol{\mathcal{T}}$ & \textcolor{black}{Stress tensor} \\
$\mathbf{u}$ & Velocity vector \\
$u_{\max}$ & Maximum axial velocity \\
$\delta$ & Interface perturbation amplitude \\
$\delta_\Gamma$ & Regularized surface delta function \\
$\varepsilon$ & Regularization parameter \\
$\dot{\gamma}$ & Shear rate \\
$\dot{\gamma}_c$ & Characteristic shear rate \\
$\kappa$ & Interface curvature \\
$\mu_{\mathrm{eff}}$ & Medium effective viscosity \\
$\mu_l$ & Newtonian thread viscosity \\
$\rho_b$ & Medium density \\
$\rho_l$ & Thread density \\
$\rho_r$ & Density ratio, $\rho_b/\rho_l$ \\
$\sigma$ & Surface tension \\
$\tau$ & Time to breakup, $t_b - t$ \\
\bottomrule
\end{tabular}
\end{multicols}
}
\end{table}

\bibliographystyle{jfm}
\bibliography{MAIN}

\end{document}